\documentclass[aps,prd,eqsecnum,11pt,showpacs,superscriptaddress,nofootinbib]{revtex4-2}
\usepackage{graphicx}
\usepackage{bm}
\usepackage{dcolumn}
\usepackage{color}
\usepackage{graphicx, tikz}
\usepackage{caption, subcaption}
\usepackage{amssymb,amsmath, mathrsfs}
\usepackage{amssymb, graphics, setspace}
\usepackage{hyperref}
\usepackage[all]{xy}
\newcommand{\be}{\begin{equation}}
\newcommand{\ee}{\end{equation}}
\newcommand{\bea}{\begin{eqnarray}}
\newcommand{\eea}{\end{eqnarray}}

\newcommand{\I}{\mathrm{i}}
\newcommand{\w}{\omega}
\begin{document}

\author{Daniel~Pe\~{n}alver}
\email{Daniel.Penalver@ific.uv.es}
\affiliation{Departamento de F\'isica Te\'orica and IFIC, Universidad de Valencia-CSIC, Av. Vicent Andr\'es Estell\'es 19, 46100 Burjassot, Spain}

\author{Marco~De~Vito}
\email{mardevi@ific.uv.es}
\affiliation{Departamento de F\'isica Te\'orica and IFIC, Universidad de Valencia-CSIC, Av. Vicent Andr\'es Estell\'es 19, 46100 Burjassot, Spain}

\author{Roberto~Balbinot}
\email{Roberto.Balbinot@bo.infn.it}
\affiliation{Dipartimento di Fisica e Astronomia dell'Universit\`a di Bologna and INFN sezione di Bologna, Via Irnerio 46, 40126 Bologna, Italy}

\author{Alessandro~Fabbri}
\email{afabbri@ific.uv.es}
\affiliation{Departamento de F\'isica Te\'orica and IFIC, Universidad de Valencia-CSIC, Av. Vicent Andr\'es Estell\'es 19, 46100 Burjassot, Spain}

\title{Sonic black holes with thick horizons in BECs}
\begin{abstract}
We consider simple one dimensional models of acoustic black holes formed by Bose-Einstein condensates where the flow is stepwise homogeneous. We shall concentrate on the case in which an extended sonic region is present, i.e. a region where the
flow velocity equals exactly the speed of sound. These models can be thought of as acoustic black holes with a thick horizon. Particular attention will be devoted to the number of created particles and to the density-density correlation functions, these latter
being at present time the experimental tool to identify Hawking like radiation in acoustic black holes.\end{abstract}

\date{\today}
\maketitle

\section{Introduction}

In General Relativity a black hole (BH) is a region of space-time which is causally disconnected from the rest of the Universe. In this region the gravitational attraction is  so strong that even light cannot escape from it and is forced to move toward the central singularity. The boundary of the BH is a null hypersurface, the event horizon.

A BH-like behaviour can be simulated by a fluid which undergoes a transition from a subsonic motion to a supersonic one \cite{unruh81, blv}. In the supersonic region sound waves are trapped inside and dragged along the flow as light in a BH. We have an acoustic BH. The transition region, where the speed of the flow equals the speed of sound, is called in analogy the sonic horizon. 

Unlike in gravity, one is not limited in this case to consider the transition from subsonic to supersonic regime to occur on an infinitely thin surface. By manipulating the external setup, one can envisage situations in which the speed of the flow equals the speed of sound over a layer of finite extension. We have a thick sonic horizon.

Among the many condensed matter systems proposed as candidates to construct an acoustic BH, the ones formed by Bose-Einstein condensates (BECs) \cite{gacz1, gacz2} are the most promising, since for them one has experimentally shown the presence of Hawking-like radiation \cite{hawking74, hawking75} by measuring its associated density correlations functions \cite{bffrc, cfrbf, jeff2016, jeff2019, jeff2021}.

In this paper we will consider simple toy models of acoustic BHs where an extended sonic region is present (preliminary results were presented in \cite{pdbf}). The models we have in mind, which are widely used in the literature \cite{rpc, mfr, lrpc, capitolo-libro}, consist of a condensate formed by gluing homogeneous regions which can be subsonic, supersonic and also sonic.

In section (\ref{hc}) we give a short introduction to the properties of homogeneous BECs in one spatial dimension, focussing the attention on the difference in the dispersion relation corresponding to the three cases analysed (subsonic, supersonic, sonic).
In section (\ref{sh}) models consisting in gluing two regions of different nature are examined and compared. Finally in section (\ref{trm}) a model in which all the three different regions are present is discussed. In all cases particular emphasis will be devoted to the density
correlation functions, which are, as already said, the basic experimental tool to study Hawking-like radiation in these settings.

\section{Homogeneous (quasi) condensates in one dimension}
\label{hc}

Using standard BEC theory  (see for example \cite{pi-st}), one decomposes the basic boson field operator $\hat\Psi$ as
\be \label{uno} \hat\Psi(t,\vec x)=\Psi_0(\vec x)\,[1+\hat \phi(t,\vec x)]\ , \ee
where the condensate is described by the macroscopic ground state wave function $\Psi_0$ which satisfies the Gross-Pitaevskii (GP) equation
\begin{equation}\label{due}
i\hbar\frac{\partial \Psi_0 }{\partial t} = \left(-\frac{\hbar^2}{2m}\vec \nabla^2 + V_{ext} + g |\Psi|^2 \right)\,\Psi_0\ . 
\end{equation}
Here $V_{ext}$ is the external potential and $g$ is the atom-atom interaction coupling.
The quantum operator $\hat \phi$ describes small fluctuations above the condensate and represents the non condensed part of the system. It satisfies the Bogoliubov-de Gennes (BdG) equation
\begin{equation}\label{tre}
i\hbar \frac{ \partial \hat \phi}{dt}= - \left( \frac{\hbar^2 \vec \nabla^2}{2m} + \frac{\hbar^2}{m}\frac{\vec \nabla \Psi_0 }{\Psi_0} \vec \nabla\right)\hat\phi +ng (\hat\phi + \hat\phi^{\dagger}).
\end{equation}
where $n=|\Psi_0|^2$ is the condensate density.

In a homogeneous 1D  (quasi) condensate the condensate wave function $\Psi_0$ can be written as
\be \label{quattro} \Psi_0=\sqrt{n}e^{ik_0x-i\omega_0t}\ , \ee
where $v=\frac{\hbar k_0}{m}$ is the flow velocity and $\hbar \w_0=\hbar^2 k_0^2/ (2m)+gn+V_{ext}$.
The fluctuation field $\hat \phi$ can be expanded as follows
\be \label{cinque}
 \hat\phi (t,x) =\sum_j  \left[ \hat a_j \phi_j (t,x) + \hat a_j^{\dagger} \varphi_j^*(t,x) \right]\ ,\end{equation}
 where $\hat a$ and $\hat a^\dagger$ are annihilation and creation operators respectively which satisfy the standard boson commutation rules $[\hat a_i, \hat a_j^\dagger] =\delta_{ij}$, and all the other commutators vanish.
The modes $\phi, \varphi$ satisfy the equations
\bea
    \label{sei}
    \left[ i (\partial_t + v \partial_x) + \frac{\xi c}{2}\partial_x^2 - \frac{c}{\xi}\right] \phi_j &=& \frac{c}{\xi} \varphi_j \\
    \label{sette}
    \left[-i (\partial_t + v \partial_x) + \frac{c\xi}{2}\partial_x^2 - \frac{c}{\xi}\right] \varphi_j &=& \frac{c}{\xi} \phi_j\ , 
\eea
where $c=\sqrt{\frac{gn}{m}}$ is the speed of sound and $\xi=\frac{\hbar}{mc}$ the healing length. These equations can be solved in terms of plane waves
\begin{equation}\label{otto}
\phi_{\omega}=D(\omega)e^{-i\omega t+ik(\omega)x},\  \varphi_{\omega}=E(\omega)e^{-i\omega t+ik(\omega)x}\ .
\end{equation}     

The frequency $\omega$ and the wave vector $k$ are related by the dispersion relation
\begin{equation}\label{nove}
(\omega-vk)^2=c^2\left(k^2+ \frac{\xi^2 k^4}{4}\right)\ .
\end{equation}     
The prefactors $D(\omega)$ and $E(\omega)$ are the normalization factors
\be\label{dieci}
D(\omega) =  \frac{\omega -v k+\frac{c\xi k^2}{2}}{\sqrt{4\pi \hbar n c\xi k^2\left| (\omega-vk) \left(\frac{dk}{d\omega}\right)^{-1} \right| }},\ee
\be\label{undici}
E(\omega) = -\frac{\omega -v k-\frac{c\xi k^2}{2}}{\sqrt{4\pi \hbar nc\xi k^2\left| (\omega-vk) \left(\frac{dk}{d\omega}\right)^{-1} \right| }}\ .
\ee  
The sign of the quantity
\be\label{dodici}
\omega-vk=\pm c\sqrt{k^2+\frac{\xi^2k^4}{4}}\equiv \Omega_{\pm}(k)\ee
 refers to the positive ($+$) or  negative ($-$)  norm branch.

At fixed $\omega$, Eq. (\ref{nove}) admits four solutions for $k(\omega)$ which vary according to the type of homogeneous condensate considered.

\subsection{Modes Analysis}
\label{ma}

We shall consider three types of homogeneous condensates, those with $|v|<c$, $|v|>c$ and $|v|=c$, corresponding respectively to subsonic, supersonic and sonic flows.


\subsubsection{Subsonic flows}

\begin{figure}[h]
    \centering
    \includegraphics[width = 0.4\textwidth]{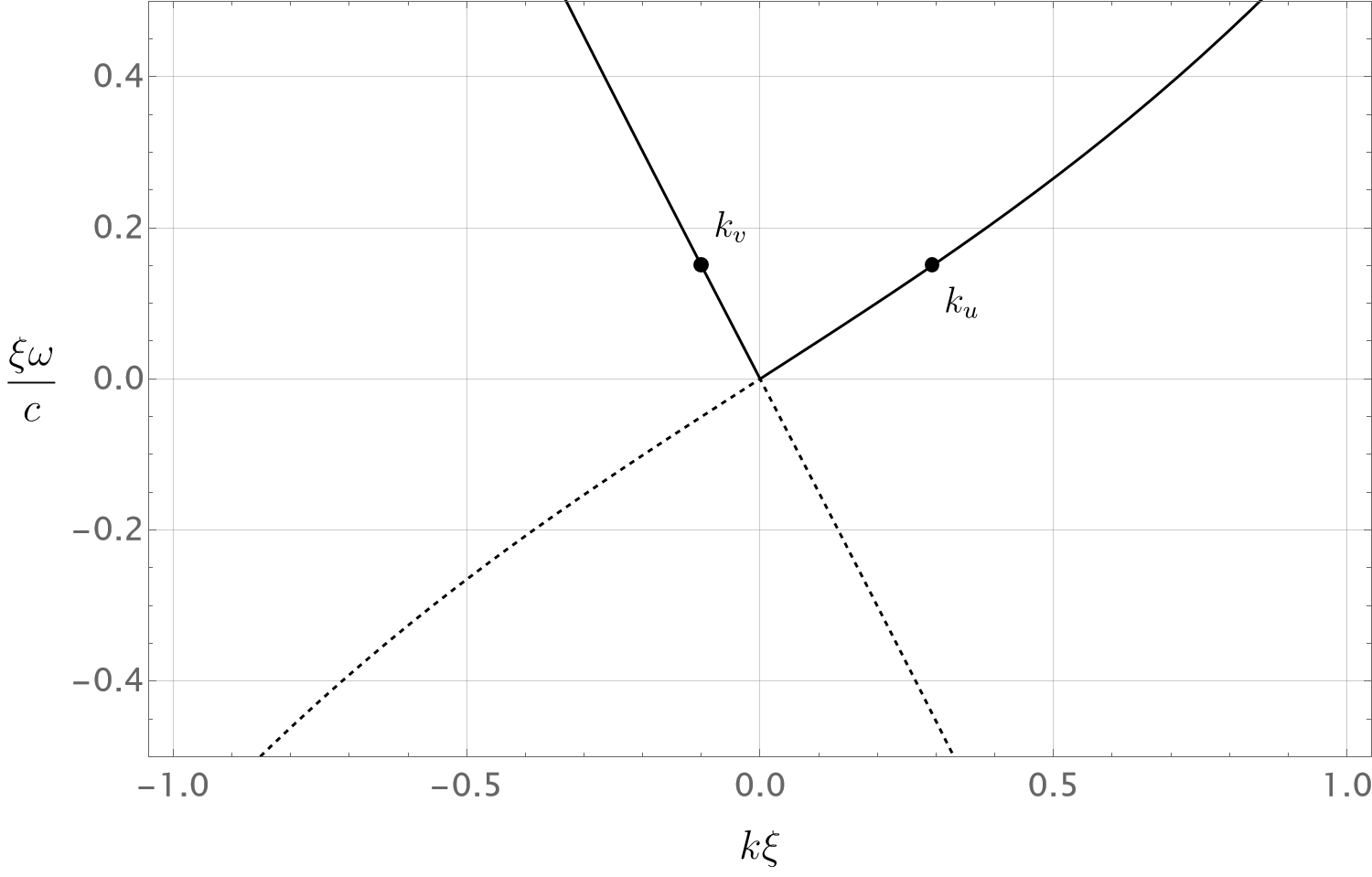}
    \caption{ Dispersion relation \eqref{nove} for a subsonic region. Solid and dashed lines represent the positive and the negative norm branch in (\ref{dodici}) respectively. For a particular (real) frequency, two solutions are real and the other two are complex conjugates. Parameters: $|\frac{v}{c}| = 0.5$.}
    \label{fig:dispersion-subsonic}
\end{figure}

We shall consider a flow directed from right to left (i.e. $v<0$). In subsonic flows the speed of the condensate is lower than the speed of sound ($|v| < c$). In this case the dispersion relation (\ref{nove}) admits two real and two complex conjugate roots (see Fig. (\ref{fig:dispersion-subsonic})). 
The two real solutions reduce, as $\xi\to 0$, to the two hydrodynamic co-current (i.e. moving to the left) $k_v=\frac{\omega}{v-c}$ and counter-current (moving to the right) $k_u=\frac{\omega}{v+c}$ modes. 
The other two solutions  are absent when $\xi=0$ and are thus completely dispersive.
We label them $k_+$ and $k_-$ according to their imaginary part's sign. 
$k_+$ defines a solution $\sim e^{ik_+x}$ that decays to $0$ as $x\to +\infty$ and diverges when $x\to -\infty$. The $k_-$ mode behaves the other way around. At low frequencies the four solutions take the form
\bea
    \label{eqn:ch4:subsonic-k-start}
    k_v &=& \frac{\w}{v - c} + \frac{1}{8}\frac{c\xi^2 \w^3}{(v-c)^4} + O(\w^5)\ , \\
    k_u &=& \frac{\w}{v + c} - \frac{1}{8}\frac{c\xi^2 \w^3}{(v+c)^4} + O(\w^5)\ ,  \\
    k_\pm &=& \pm 2\I \frac{\sqrt{c^2 - v^2}}{c\xi} + \frac{v\w}{c^2 - v^2} \pm \frac{\I}{4} \frac{(c^2 + 2v^2) c\xi\w^2}{(c^2-v^2)^{5/2}} - \frac{1}{2}\frac{(c^2 + v^2)vc^2\xi^2\w^3}{(c^2-v^2)^4} + O(\w^4)\ . \ \ \ \ \ 
    \label{eqn:ch4:subsonic-k-end}
\eea

The normalization coefficients $D$ and $E$ in this limit are given in Appendix \ref{appendixA}.

\subsubsection{Supersonic flows}

\begin{figure}[h]
    \centering
    \includegraphics[width = 0.4\textwidth]{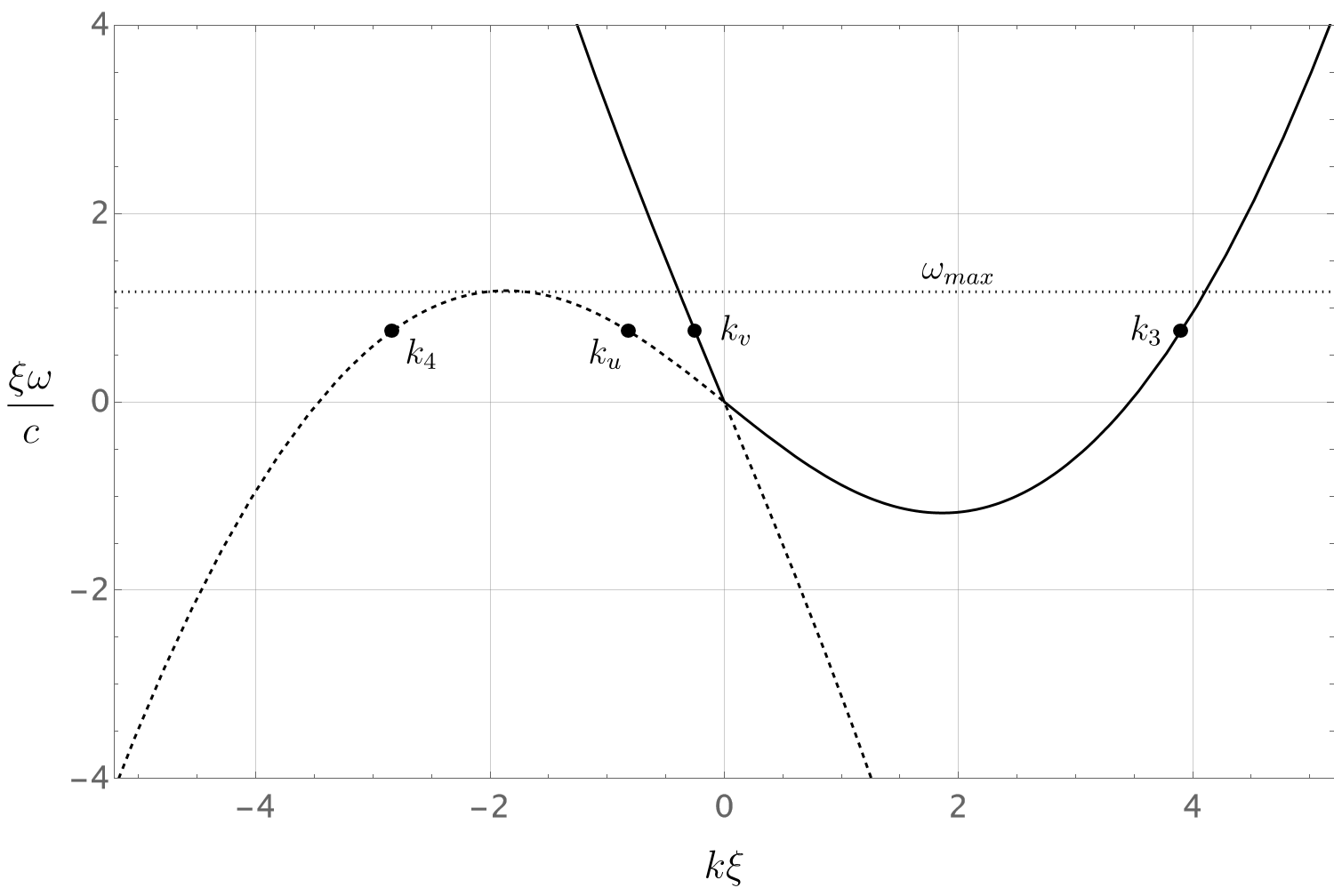}
    \caption{Dispersion relation \eqref{nove} for a supersonic region. Solid and dashed lines represent the positive and the negative branch of the square root in (\ref{dodici}). For frequencies $\w < \w_{max}$, all four roots are real and two of them belong to the negative norm branch. For frequencies $\w > \w_{max}$, only two of them are real as in the subsonic case. Parameters: $|\frac{v}{c}| = 2$.}
 \label{fig:dispersion-supersonic}
\end{figure}

In supersonic flows the speed of the condensate is greater than the speed of sound ($|v| > c$). 
The plot of the dispersion relation (\ref{nove}) of 
Fig. (\ref{fig:dispersion-supersonic})
 shows a clear difference from the subsonic case. 
Indeed, there exists a maximal frequency $\w_{max}\sim \frac{1}{\xi}$  \cite{mapa}
below which all four solutions of the dispersion relation (\ref{nove}) are real. Due to the supersonic character of the flow both the hydrodynamic modes (i.e. those reducing to $k_v=\frac{\omega}{v-c}, k_u=\frac{\omega}{v+c}$ as $\xi\to 0$) move to the left ($k_u$, that was counter-current right-moving in subsonic flows, is now dragged by the fluid to the left). The two additional real modes, that we denote by $k_3, k_4$, move both to the right (their group velocity is positive, $\frac{d\omega}{dk}>0$). These are supersonic modes arising because of  dispersion. As for the subsonic $k_\pm$ modes, they disappear as $\xi\to 0$.


We also remark, by looking at  Fig. 2,  that two of these modes, $k_u$ and $k_4$, belong to the negative branch of the dispersion relation (\ref{nove}) and are thus associated to negative-norm modes. 

For $\w > \w_{max}$, instead, we have the behaviour seen in the subsonic case with only two real solutions, corresponding to co-current $k_v$ and counter-current $k_3$ modes.

For small $\omega$,  the wavevectors $k_v,k_u,k_3,k_4$ take the form


\bea
    \label{eqn:ch4:supersonic-k-start} k_v &=& \frac{\omega}{v - c} + \frac{1}{8}\frac{c \xi^2 \omega^3}{(v - c)^4} + O(\w^5)\ , \\
    k_u &=& \frac{\omega}{v + c} - \frac{1}{8}\frac{c \xi^2 \omega^3}{(v + c)^4} + O(\w^5)\ , \\
    k_{3,4} &=& \pm 2 \frac{\sqrt{v^2 - c^2}}{c \xi} - \frac{v\w}{v^2 - c^2} \mp \frac{1}{4}\frac{(c^2 + 2v^2) c\xi\w^2}{(v^2 - c^2)^{5/2}} - \frac{1}{2}\frac{(c^2 + v^2) v c^2\xi^2\w^3}{(v^2 - c^2)^4} + O(\w^4)\ \ \ \ 
    \label{eqn:ch4:supersonic-k-end} 
\eea
and the corresponding expressions for $D$ and $E$ are given in Appendix \ref{appendixA}.



\subsubsection{Sonic flows}

\begin{figure}[h]
   \centering
   \includegraphics[width = 0.4\textwidth]{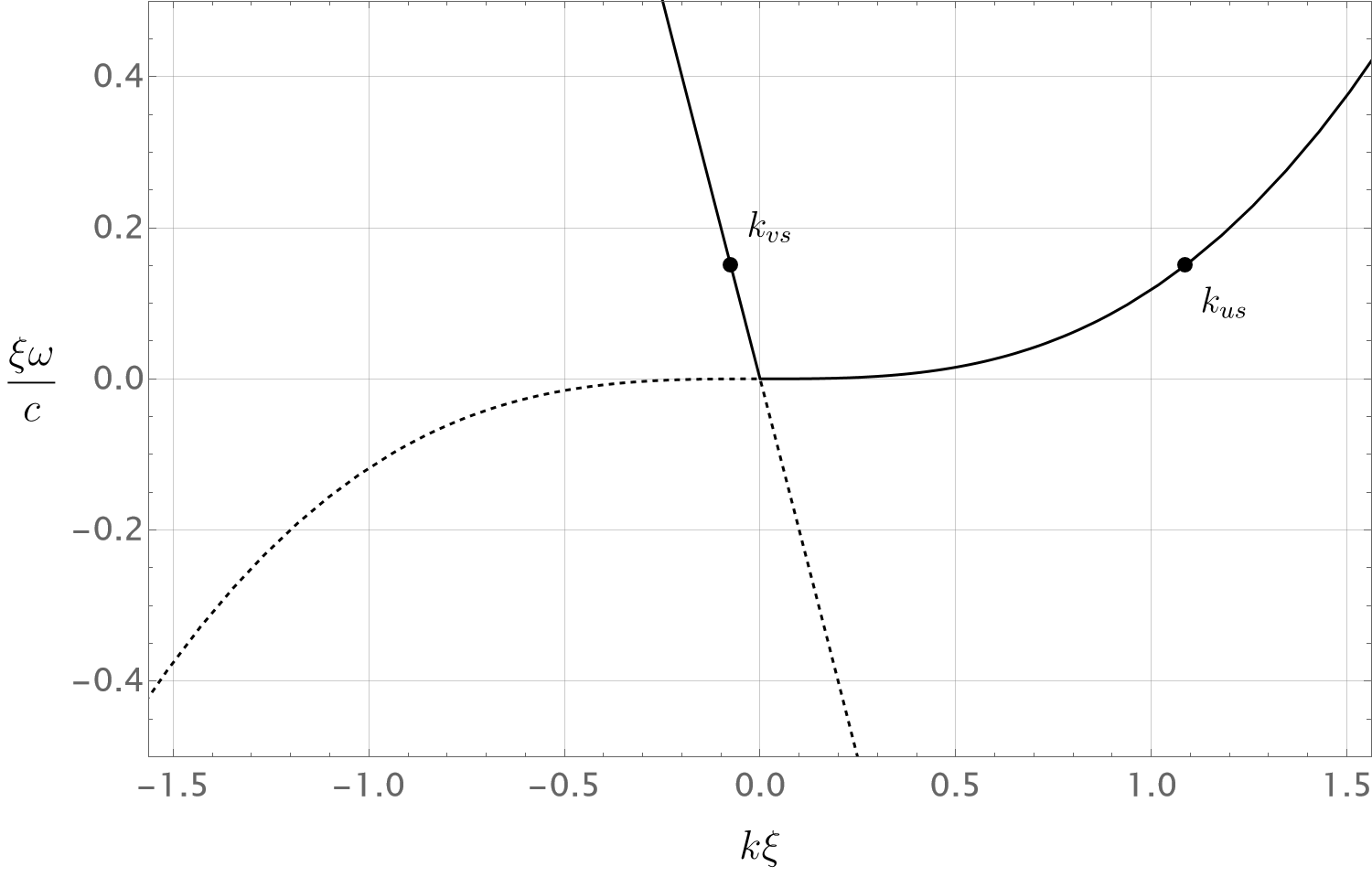}
  \caption{Dispersion relation \eqref{nove} for a sonic region. Solid and dashed lines represent the positive and negative branch of the square root in (\ref{dodici}). Parameters: $|\frac{v}{c}| = 1$.}
\label{fig:dispersion-sonic}
\end{figure}

For a sonic flow 
the speed of the condensate equals the speed of sound ($|v| = c$). In this limit the subsonic $k_u,k_\pm$  (Eqs. (2.14), (2.15)) and supersonic $k_u, k_{3,4}$ (Eqs. (2.17), (2.18)) solutions diverge. The plot of the solutions of (\ref{nove}) for this case, given in Fig. (\ref{fig:dispersion-sonic}), shows a qualitative behaviour similar to the subsonic case: one co-current ($k_{vs}$) and one counter-current ($k_{us}$) mode solutions, both of positive norm.  The two additional solutions to (\ref{nove}), $k_{\pm s}$, are complex conjugates. Out of the four solutions only one, $k_v$, survives in the hydrodynamical $\xi\to 0$ limit, and
is the $v\to -c$ limit of $k_v=\frac{\omega}{v-c}$ seen in the previous cases. The outgoing solution $k_{us}$ and $k_{\pm s}$ are totally dispersive.
In the small $\omega$ limit we have 
 

\bea
    k_v &=& -\frac{\omega}{2 c} + \frac{1}{384}\frac{\xi^2 \omega^3}{c^3} + O(\omega^5)\ ,  \\
    k_u &=& \frac{1}{\xi}\left[ 2 \left(\frac{\xi \w}{c}\right)^{1/3} + \frac{1}{6}\left(\frac{\xi \w}{c}\right) - \frac{1}{36}\left(\frac{\xi \w}{c}\right)^{5/3} + \frac{5}{648} \left(\frac{\xi \w}{c}\right)^{7/3} \right] +O(\w^{3})\ , \label{tutud} \\
    k_\pm &=& \frac{1}{\xi}\left[ (-1\pm\I\sqrt{3}) \left(\frac{\xi \w}{c}\right)^{1/3} + \frac{1}{6}\left(\frac{\xi \w}{c}\right) - \frac{1 \pm \I\sqrt{3}}{72}\left(\frac{\xi \w}{c}\right)^{5/3} \right] + O(\w^{7/3})\ . 
\eea
The corresponding $D$ and $E$ normalization coefficients in this limit are given in Appendix \ref{appendixA}.\footnote{BEC phase fluctuations behave in the hydrodynamical approximation like a  massless scalar field propagating in an effective metric of the form $ds^2= \frac{n}{mc}[ -(c^2-v^2)dt^2 -2vdtdx +dx^2+dy^2 +dz^2]$, where the flow is along the $x$ direction. For  a sonic flow
the first term vanishes and outgoing null (i.e. $ds^2=0$) longitudinal (i.e. $y,z$ constant) rays obey the equation $x=$const.: they are ``frozen'' inside the sonic region. It should be stressed, however, that the model presented here is completely dispersive and we are outside the hydrodynamical regime as indicated by eq. (\ref{tutud}).}

\section{Stepwise homogeneous condensates}
\label{sh}

We will consider a very idealised setting consisting of a 1D BEC formed by homogeneous regions connected by steplike discontinuities. This kind of models is extensively used in the literature and we will refer mostly to the formalism presented in the pedagogical review of Ref. \cite{capitolo-libro}.

To simplify the discussion, we assume the condensate density $n$ and the flow velocity $v$ (from right to left, i.e. $v<0$) to be the same in all regions. On the other hand, the external potential $V_{ext}$ and the coupling constant $g$ have different values in each sector, but satisfy the relation  
\be \label{quno} V_{ext}^l+g^ln=V_{ext}^r+g^rn\ , \ee
where the superscripts `l' and `r' refer to the left and to the right respectively of a steplike discontinuity gluing two neighbouring zones. Changing $g$ from one sector to the other allows the speed of sound $c$ to have different values in different regions, since $mc^2=gn$. So one can have a region where the flow is subsonic ($|v|<c$), another where it is supersonic ($|v|>c$) and even a sonic ($|v|=c$) one.
Due to the condition (\ref{quno}) the wave function 
\be \label{qdue} \Psi_0=\sqrt{n}e^{ik_0x-i\omega_0t}\ee
is a plane wave solution of the GP equation (\ref{due}) everywhere and at all times.
Concerning the BdG equation for the fluctuations (\ref{tre}), the modes $\phi$ and $\varphi$ have to satisfy the following matching conditions across each different discontinuity
\begin{equation}\label{qtre}
[\phi]=0,\, [\phi']=0,\, [\varphi]=0,\, [\varphi']=0,\ee
where $[f(x)]=\lim_{\epsilon\to 0} [f(x+\epsilon)-f(x-\epsilon)]$ and a prime `\ ${}^\prime$\ ' means derivative with respect to $x$.
Coming back to the dispersion relation, at fixed $\omega$ this is a fourth order equation in $k$, admitting four solutions $k_i(\omega)$. Hence $\phi$ is a linear combination of four plane waves
\be\label{qquattro}\phi_\omega(x,t)=e^{-i\omega t}\sum_{i=1}^4A_{ij}(\omega)D_{ij}(\omega)e^{ik_{ij}(\omega)x}\ , \ee
where the $A_{ij}(\omega)$ represent the amplitude of each plane wave, $D_{ij}(\omega)$ is the corresponding normalisation factor and the index $j$ labels the different regions. For $\varphi$ one has a similar expansion with the substitution $D(\omega)\to E(\omega)$.

The four matching conditions (\ref{qtre}) allow to establish a linear relation between the four amplitudes on the left of each discontinuity and the ones on the right
\be \label{qcinque} A_{il}=M_{ij}A_{jr}\ .\ee
$M$ is a $4\times 4$ matrix called the matching matrix.

\subsection{Two regions model: subsonic-supersonic}
\label{susu}

This model has been largely discussed in the literature and since for us this configuration and its features serve only as reference to compare with the analogous ones of other models, we rapidly go through the theoretical construction following Ref. \cite{capitolo-libro}.
The model consists of two semi-infinite homogeneous regions glued along a steplike discontinuity located at $x=0$. The right region ($x>0$) `r' is subsonic, while the left one `l', $x<0$, is supersonic. This latter represents the sonic BH, bounded by the sonic horizon at $x=0$.

According to our previous modes analysis, in the subsonic region we have two propagating modes with real momentum, $k_u$ and $k_v$, and two others with complex $k$, namely $k_+$ and $k_- $. We have to exclude the $k_-$ solution as its corresponding
mode diverges as $x \to  +\infty$. So we have to set the corresponding amplitude to zero, i.e. $A_{-r} =0$. In the supersonic region for $\omega<\omega_{max}$ we have four propagating modes $k_u, k_v, k_3$ and $k_4$.
\footnote{We will not discuss the $\omega>\omega_{max}$ case since it is trivial, involving just the scattering of two incoming modes into two outgoing modes without leading to particles creation we are interested in.}
The matching matrix $M$ can then be computed as $M=W_l^{-1} W_r$, where
\be
\label{qsei}
W_{l}=\left(
     \begin{array}{cccc}
       D_{vl} & D_{ul} & D_{3l} & D_{4l}\\
       ik_{vl}D_{vl} & ik_{ul}D_{ul} & ik_{3l}D_{3l} & ik_{4l}D_{4l} \\
       E_{vl} & E_{ul} & E_{3l} & E_{4l} \\
       ik_{vl}E_{vl} & ik_{ul}E_{ul} & ik_{3l}E_{3l} & ik_{4l}E_{4l}  \\
\end{array}\right)\,, 
\end{equation}
and
\be
\label{qsette}
W_{r}=\left(
     \begin{array}{cccc}
       D_{vr} & D_{ur} & D_{+r} & D_{-r}\\
       ik_{vr}D_{vr} & ik_{ur}D_{ur} & ik_{+r}D_{+rl} & ik_{-r}D_{-r} \\
       E_{vr} & E_{ur} & E_{+r} & E_{-r} \\
       ik_{vr}E_{vr} & ik_{ur}E_{ur} & ik_{+r}E_{+r} & ik_{-r}E_{-r}  \\
\end{array}\right)\,. 
\end{equation}

\begin{figure}[h]
   \centering
   \includegraphics[width = 0.3\textwidth]{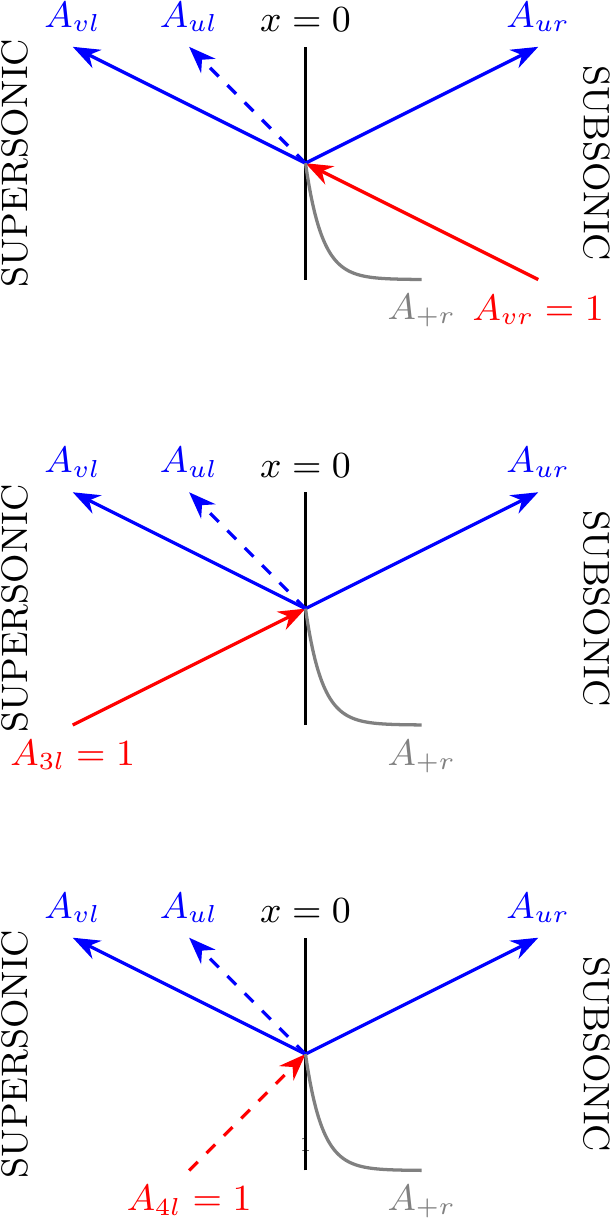}
  \caption{`in' modes for the subsonic-supersonic model.}
\label{in-modes}
\end{figure}

\begin{figure}[h]
   \centering
   \includegraphics[width = 0.3\textwidth]{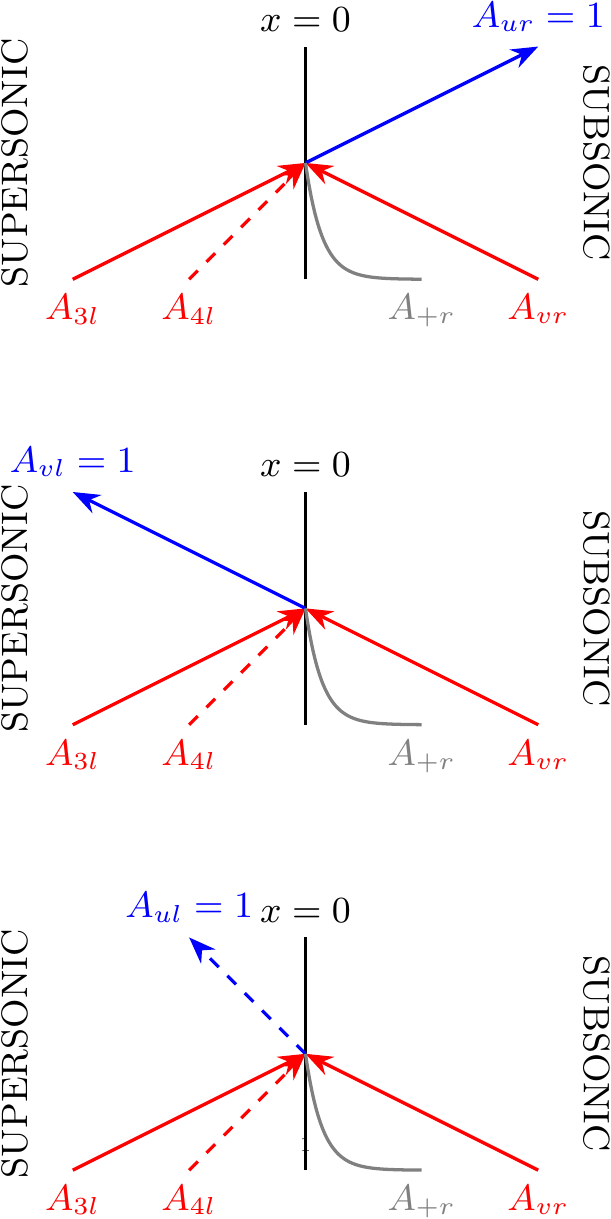}
  \caption{`out' modes for the subsonic-supersonic model.}
\label{out-modes}
\end{figure}

One then proceeds to construct the `in' scattering basis from the three incoming (i.e. directed from infinity towards the discontinuity) modes with momenta $k_{3l}$ and $k_{4l}$ (coming from $-\infty$) and $k_{vr}$ (coming from $+ \infty$) which we denote as
$\phi_{3l}^{in}, \phi_{4l}^{in}, \phi_{vr}^{in}$ respectively. The definition of these modes is sketched in Fig. (\ref{in-modes}).
Similarly for the `out'  basis with momenta $k_{ur}, k_{ul}, k_{vl}$, namely $\phi_{ur}^{out}, \phi_{ul}^{out}, \phi_{vl}^{out}$ as depicted in Fig. (\ref{out-modes}).
The field operator $\hat \phi$ can be expanded in terms of the `in' basis or equivalently in terms of the `out' basis
\bea
    \hat \phi &=& 
    \int \bigg(\phi^{in}_{vr} \hat a^{in}_{vr}(\w) + \phi^{in}_{3l} \hat a^{in}_{3l}(\w) + \phi^{in}_{4l} \hat a^{in\dagger}_{4l}(\w) 
    +  \varphi^{in*}_{vr} \hat a^{in\dagger}_{vr}(\w) +  \varphi^{in*}_{3l} \hat a^{in\dagger}_{3l}(\w) + \varphi^{in*}_{4l} \hat a^{in}_{4l}(\w)\bigg)d\omega \nonumber \nonumber \\
    &=& \int \bigg(\phi^{out}_{ur} \hat a^{out}_{ur}(\w) + \phi^{out}_{vl} \hat a^{out}_{vl}(\w) + \phi^{out}_{ul} \hat a^{out\dagger}_{ul}(\w) 
    +  \varphi^{out*}_{ur} \hat a^{out\dagger}_{ur}(\w) + \varphi^{out*}_{vl} \hat a^{out\dagger}_{vl}(\w) \nonumber \\ &+& \varphi^{out*}_{ul} \hat a^{out}_{ul}(\w)   \bigg)d \w .
\label{eqn:ch4:in-out-expansion}
\eea
 
Note the appearance of a creation operator in the third term of the first line and in the second line. This because the corresponding momenta $k_{4l}$ and $k_{ul}$ belong to the negative branch of the dispersion relation and the associated modes have negative norm.
The `in' and `out' basis are related by a $3\times3$ scattering matrix $S$

\begin{eqnarray}\label{eq:outinaa}
  \phi_\w^{v,in}&=& S_{vl,vr} \phi_\w^{v,out}+S_{ur,vr} \phi_\w^{ur,out}+S_{ul,vl} \phi_\w^{ul,out}\ , \\
  \phi_\w^{3,in} &=& S_{vl,3l} \phi_\w^{v,out}+S_{ur,3l} \phi_\w^{ur,out}+S_{ul,3l} \phi_\w^{ul,out}\ , \\
  \phi_\w^{4,in} &=& S_{vl,4l} \phi_\w^{v,out}+S_{ur,4l} \phi_\w^{ur,out}+S_{ul,4l} \phi_\w^{ul,out}\ ,
\end{eqnarray}
where the notation introduced for the $S$ matrix elements is the one of Ref. \cite{capitolo-libro} where the first index refers to the outgoing channel and the second index to the incoming one. The (leading order in $\omega$) elements of the $S$ matrix are given in Appendix \ref{appendixB}.
The corresponding Bogoliubov transformation between creation and annihilation operators of the two bases are

\begin{equation}
\label{Ssupsup}
     \left( \begin{array}{c}
       \hat a_{\omega}^{v,out} \\
       \hat a_{\omega}^{ur,out}  \\
       \hat a_{\omega}^{ul,out\dagger} \\
           \end{array} \right)
   = \left(
     \begin{array}{cccc}
       S_{vl,vr} & S_{vl,3l} & S_{vl,4l} \\
       S_{ur,vr} & S_{ur,3l} & S_{ur,4l} \\
       S_{ul,vr} & S_{ul,3l} & S_{ul,4l} \\
       \end{array}\right) \left(
                \begin{array}{c}
                  \hat a_{\omega}^{v,in}  \\
                  \hat a_{\omega}^{3in} \\
                  a_{\omega}^{4in\dagger}\\
                                  \end{array}
              \right) \ .
\end{equation}

Since this transformation mixes creation and annihilation operators, the Bogoliubov transformation is not trivial, so the $| in,0\rangle$ vacuum does not coincide with the $| out,0\rangle$ vacuum. We have particles (phonons) creation due to a parametric process taking place at the discontinuity surface (the horizon).

For the number of created particles at frequency $\omega$ we have
\begin{eqnarray}\label{sund}
n_\w^{u,r}&=&\langle 0,in|\hat a^{ur,out\dagger}_\w\hat a^{ur,out}_\w|0,in\rangle =|S_{ur,4l}|^2\ , \label{uuuu}\\
n_\w^{u,l}&=&\langle 0,in|\hat a^{ul,out\dagger}_\w\hat a^{ul,out}_\w|0,in\rangle=|S_{ul,vr}|^2+|S_{ul,3l}|^2\ , \nonumber \\
n_\w^{v,l}&=&\langle 0,in|\hat a^{v,out\dagger}_\w\hat a^{v,out}_\w|0,in\rangle=|S_{vl,4l}|^2\ . \nonumber \end{eqnarray}
Note that these numbers satisfy the relation
\begin{equation}\label{sdod}
n_\w^{u,l}=|S_{ul,vr}|^2+|S_{ul,3l}|^2=|S_{ur,4l}|^2+|S_{vl,4l}|^2=n_\w^{u,r}+n_\w^{v,l}\ .
\end{equation}

\begin{figure}[h]
   \begin{subfigure}{0.8\textwidth}
   \includegraphics[width = 0.5\textwidth]{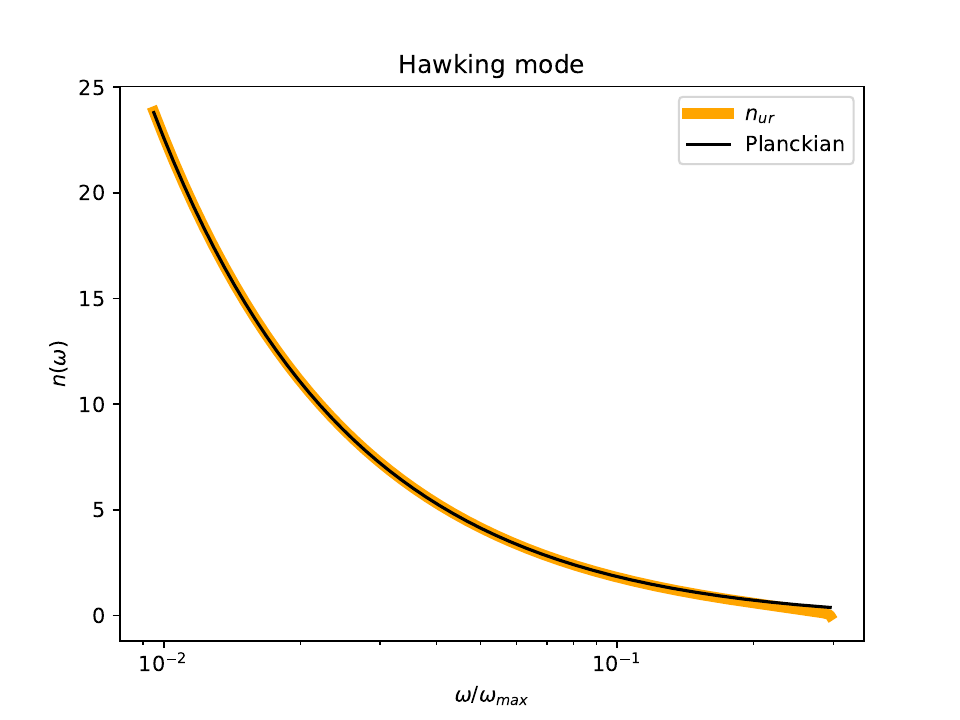}
  \caption{}
\end{subfigure}
\hfill
\begin{subfigure}{0.8\textwidth}
   \includegraphics[width = 0.5\textwidth]{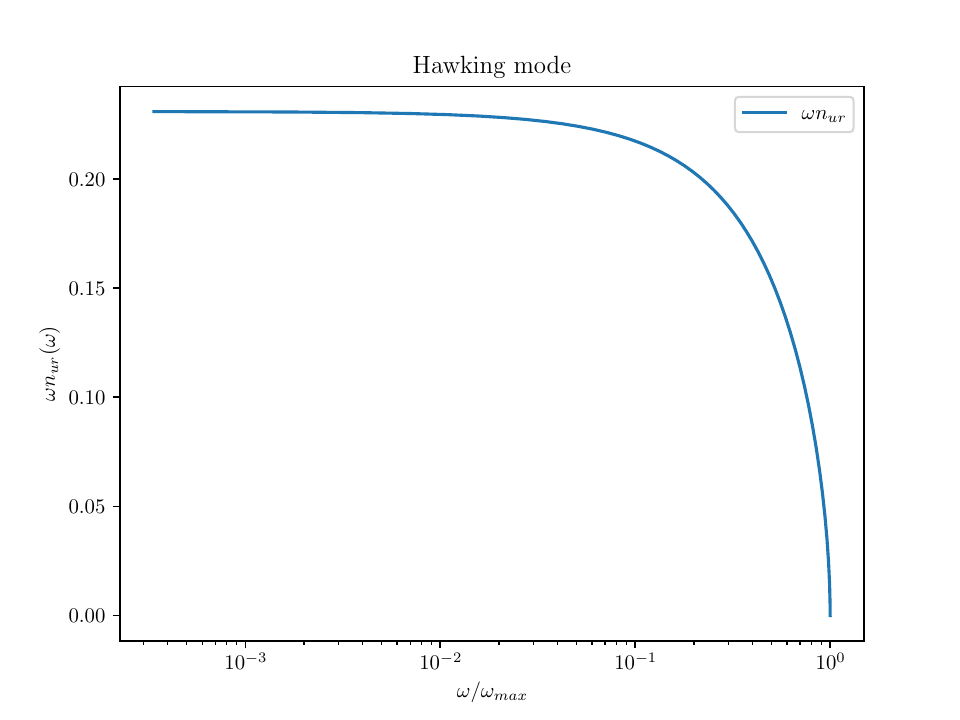}
  \caption{}
\end{subfigure}
\caption{a) Comparison between $n_{ur}$ and a Planckian distribution with $T=T_H$ given in  Eq. (\ref{htt}).
b) Plot of $\omega n_{ur}(\omega)$. Here and in the figures that follow we chose the values $c_r = 2, v=-1, c_l = 0.5, m = 1, n = 1/(4\pi), \hbar=1$.}
\label{planck0-nur-ss}
\end{figure}

\begin{figure}[h]
   \begin{subfigure}{0.8\textwidth}
      \includegraphics[width = 0.5\textwidth]{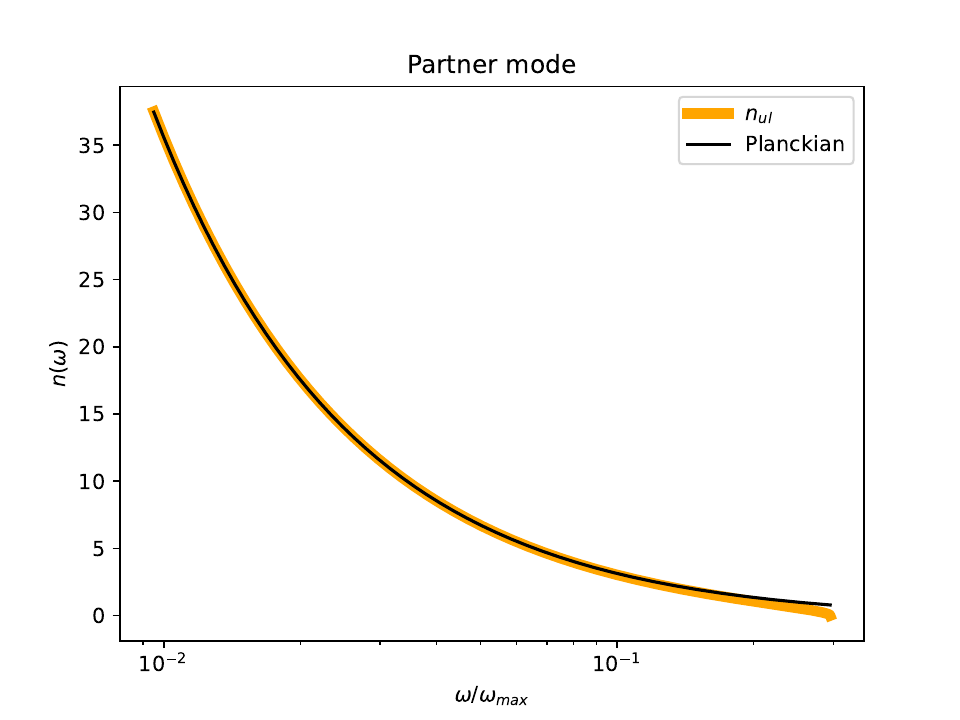}
  \caption{}
\end{subfigure}
\hfill
   \begin{subfigure}{0.8\textwidth}
      \includegraphics[width = 0.5\textwidth]{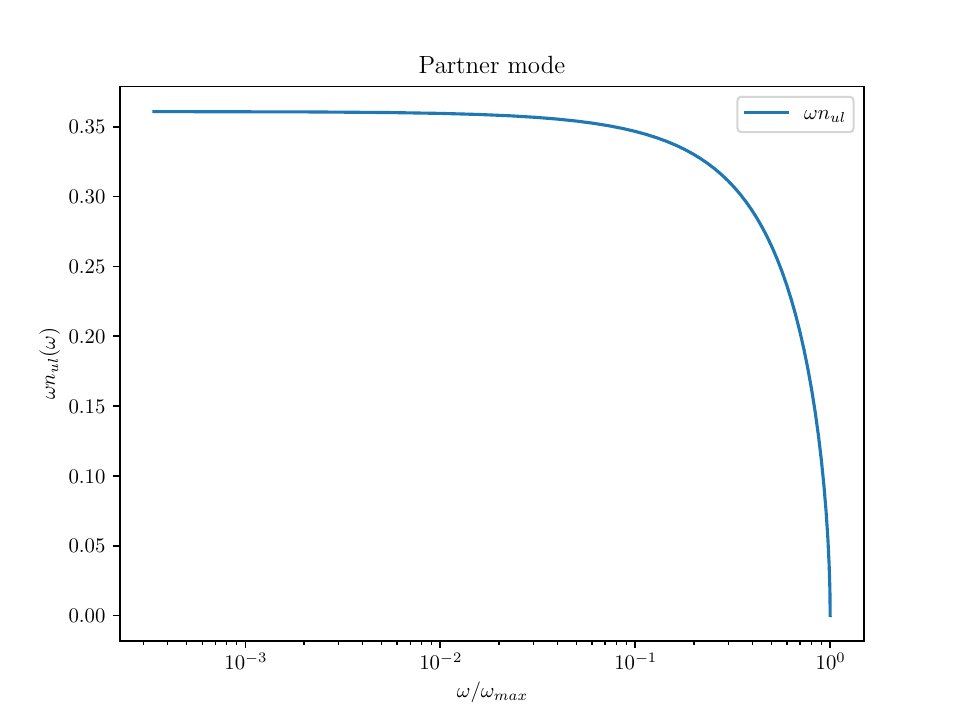}
  \caption{}
\end{subfigure}
\caption{a) Comparison between $n_{ul}$ and a Planckian distribution  with $T=T_P$ given in Eq. (\ref{ptt}).
b) Plot of $\omega n_{ul}(\omega)$.}
\label{planck2-nul-ss}
\end{figure}

\begin{figure}[h]
  \begin{subfigure}{0.8\textwidth}
     \includegraphics[width = 0.5\textwidth]{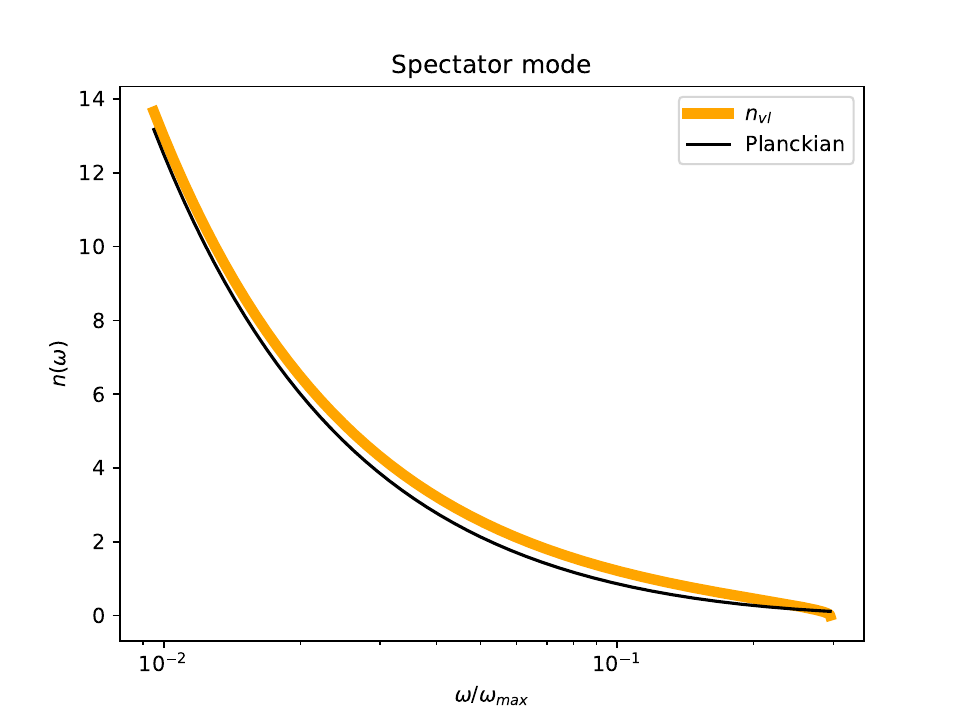}
  \caption{}
\end{subfigure}
\begin{subfigure}{0.8\textwidth}   
      \includegraphics[width = 0.5\textwidth]{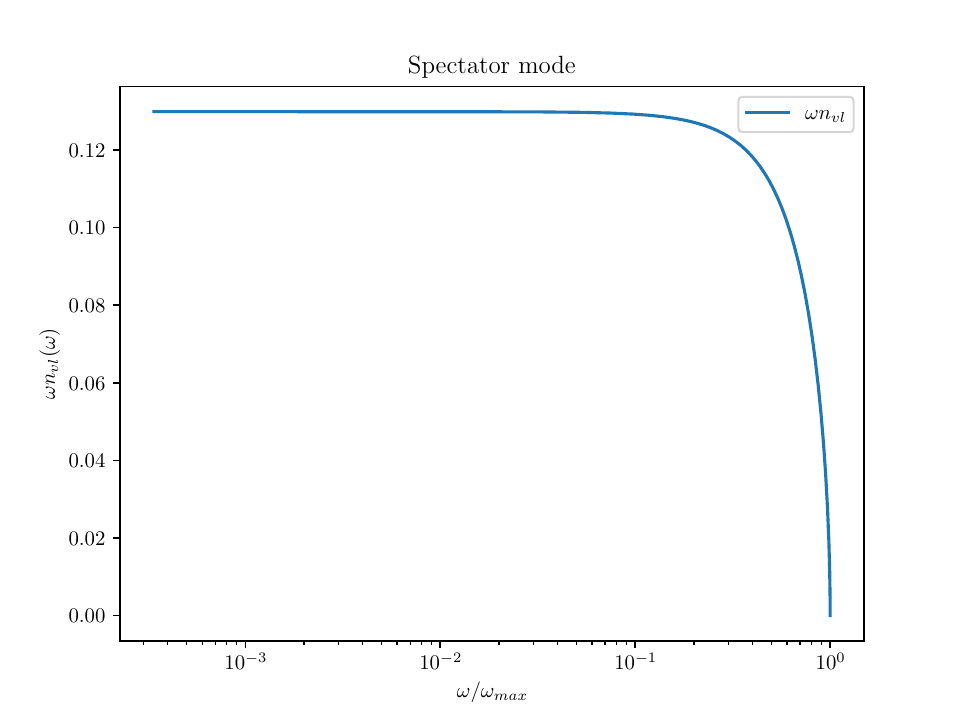}
  \caption{}
\end{subfigure}
\hfill
\caption{a) Comparison between $n_{vl}$ and a Planckian distribution with $T=T_S$ given in Eq. (\ref{stt}).
b) Plot of $\omega n_{vl}(\omega)$.}
\label{planck1-nvl-ss}
\end{figure}

The corresponding particles are named in the literature as `Hawking particle', the one in the exterior region with momentum $k_{ur}$, `partner' the one inside the horizon with momentum $k_{ul}$ and `spectator' the one also inside the horizon with momentum $k_{vl}$.
Since all these numbers show at low frequency a $\frac{1}{\omega}$ behaviour, one can associate to their emission an effective temperature characteristic of a Bose distribution according to
\begin{equation}\label{td}
n_T(\w)=\frac{1}{e^{\frac{k_BT}{\hbar \w}}-1}\simeq \frac{k_B T}{\hbar \w} +...\ .
\end{equation}
From the relevant $S$ matrix elements we have for the Hawking particle
\begin{equation}\label{htt}
T_H=\frac{\hbar}{k_B} \frac{(c_r+v)}{(c_r-v)}\frac{(v^2-c_l^2)^{3/2}}{(c_r^2-c_l^2)}
\frac{2c_r}{c_l\xi_l }\ ,
\end{equation}
for the partners
\begin{equation}\label{ptt}
T_P= \frac{\hbar}{k_B} \frac{(c_r+v)}{(c_r-v)}\frac{(v^2-c_l^2)^{3/2}}{(c_r-c_l)}
\frac{(c_r+c_l)}{2c_l^2\xi_l }\ ,
\end{equation}
and for the spectators
\begin{equation}\label{stt}
T_S= \frac{\hbar}{k_B} \frac{(c_r+v)}{(c_r-v)}\frac{(v^2-c_l^2)^{3/2}}{(c_r+c_l)}
\frac{(c_r-c_l)}{2c_l^2\xi_l }\ .
\end{equation}
Comparing the numerically evaluated number of Hawking particles with a thermal distribution at $T_H$, we see 
from Fig. (\ref{planck0-nur-ss})
that the agreement is quite good for $\omega<\omega_{max}$, while for $\omega>\omega_{max}$ the emission ceases as expected, since there are
no negative norm modes.
A similar comparison is made for the partner and spectators in Figs. (\ref{planck2-nul-ss}), (\ref{planck1-nvl-ss}). 

\begin{figure}[h]
    \centering
    \includegraphics[width = 0.5\textwidth]{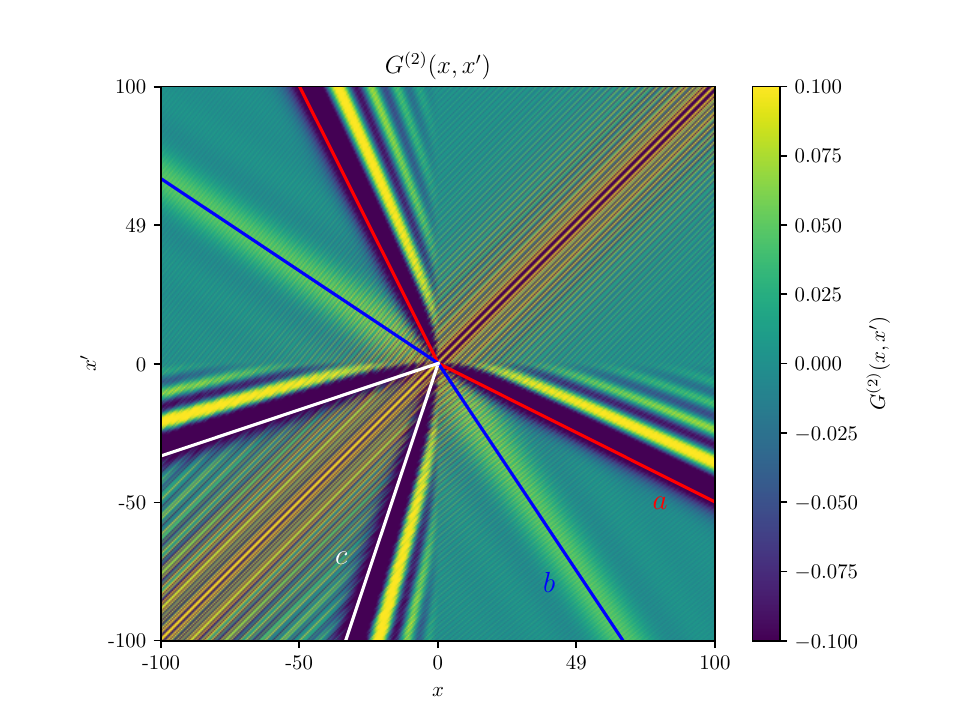}
    \caption{Two-point density correlation function for the subsonic-supersonic model. The sonic horizon is located at $x = 0$. Feature a) is given by the Hawking particles - partners correlations, b) between Hawking particles and spectators and c) between partners and spectators.
    }
    \label{fig:ch4:twopoint-sub-super}
\end{figure}

As it is well known, Hawking like emission from an acoustic BH formed by a BEC has been indeed (indirectly) observed looking at the density-density correlation function in particular at its characteristic peaks.
In the density phase representation, one can write the field operator for the fluctuations as
\begin{equation}
    \hat \phi = \frac{\hat n}{2n} + \I \hat \theta\implies \hat n = n (\hat \phi + \hat \phi^\dagger)
\label{eqn:nPhi}
\end{equation}
and the relevant (one-time) correlation function for the density operator $\hat n$ is defined as
\begin{equation}
    G^{(2)}(t;x,x') = \lim_{t^\prime \to t}\frac{1}{2n^2} \langle 0,in| [\hat n(t,x) \hat n(t^\prime,x') + \hat n (t^\prime,x') \hat n(t,x)] |0,in\rangle\ , 
\end{equation}
 where
\bea
\hat n(t,x)&=&n\int_{0}^{\omega_{max}}\Big[\hat a_{\omega}^{v,out}(\phi_\w^{v,out}+\varphi_\w^{v,out})+
\hat a_{\omega}^{ur,out}(\phi_\w^{ur,out}+\varphi_\w^{ur,out})\nonumber \\ &+& \hat a_{\omega}^{ul,out\dagger}(\phi_\w^{ul,out}+\varphi_\w^{ul,out})+h.c.\Big] d\omega\ .
\eea
$G^{(2)}$  exhibits correlations between Hawking particles and partners visible as a main peak (feature a) in Fig. (\ref{fig:ch4:twopoint-sub-super})) and also correlations between Hawking particles and spectators (feature b) in Fig. (\ref{fig:ch4:twopoint-sub-super})) when $x$ and $x’$ lie on opposite sides with respect to the
horizon, or correlations between partners and spectators if $x$ and $x'$ are both inside the horizon (feature c) in Fig. (\ref{fig:ch4:twopoint-sub-super})). 
As done in Ref. \cite{rpc}, we can give an approximate analytic form for feature a) by considering the Hawking quanta-partner (HP) correlator (we fix, for instance, $x$ outside and $x'$ inside the horizon)
\be \label{hpr} G^{(2)}_{HP}= Re \int_{0}^{\omega_{max}}d\omega S_{ur,4l}(D+E)_{ur}S_{ul,4l}^*(D+E)_{ul}e^{i(k_{ur}x-k_{ul}x')}\ee 
and using the leading order low-frequency results given in Subsection (\ref{ma}), Appendix (\ref{appendixA}) and Appendix (\ref{appendixB}).  This gives
\be \label{hpp} G^{(2)}_{HP}\sim - \frac{(v^2-c_l^2)^{3/2}}{c_l(c_l-c_r)(c_r-v)(c_l+v)}\frac{\sin \omega_{max}(\frac{x}{v+c_r}-\frac{x'}{v+c_l})}{(\frac{x}{v+c_r}-\frac{x'}{v+c_l})}\ .\ee
The (negative) peak is along the half-line $\frac{x}{v+c_r}=\frac{x'}{v+c_l}$, and represents the equal time correlation between pairs of $ur$ and $ul$ phonons created on both sides of the horizon and propagating away from it with, respectively, velocities $v+c_r$ and $v+c_l$. Away from the peak  the HP correlator (\ref{hpp}) presents damped oscillations. This is due to dispersion, i.e. to the existence of a maximal frequency $\omega_{max}$.
A similar analysis can be done for feature b), the Hawking quanta-spectator (HS) correlator
\be \label{hsr} G^{(2)}_{HS}= Re \int_{0}^{\omega_{max}}d\omega S_{ur,4l}(D+E)_{ur}S_{vl,4l}^*(D+E)_{vl}e^{i(k_{ur}x-k_{vl}x')}\ , \ee 
which can be approximated as
\be \label{hsp} G^{(2)}_{HS}\sim  \frac{(v^2-c_l^2)^{3/2}}{c_l(c_l+c_r)(c_r-v)(c_l-v)}\frac{\sin \omega_{max}(\frac{x}{v+c_r}-\frac{x'}{v-c_l})}{(\frac{x}{v+c_r}-\frac{x'}{v-c_l})}\ .\ee
In this case the peak, located along the half-line $\frac{x}{v+c_r}=\frac{x'}{v-c_l}$, is positive and its absolute value is always smaller than the HP peak, as we can see from Fig. (\ref{fig:ch4:twopoint-sub-super}).  This analysis can be reapeated also for feature c), correlations between partners and spectators, when both points are inside the horizon: this correlator presents a negative peak which  (in absolute value) is also always smaller than the HP one. 

In the next section we will see how all these features change dramatically as we replace the subsonic region of the model by a sonic one.

\subsection{Two regions model: sonic-supersonic}
\label{ss}

In this model we replace the subsonic $x>0$ region of the previous one by a sonic region, i.e. $c_l < c_r=|v|$ .
Apparently nothing should change, since in the sonic region the four solutions of the dispersion relation have the same characteristics as in a subsonic flow. We have one incoming (from $x=+\infty$) mode with real momentum $k_{vr}$ and one
outgoing (towards $x = +\infty$) with real momentum $k_{ur}$. Furthermore there are two complex solutions $k_{+r}$ and $k_{-r}$. The mode associated to this latter has vanishing amplitude since it diverges at $x=+\infty$.
The mathematical construction of the matching matrix $M$, the `in' and `out' basis and the scattering $3\times3$ matrix $S$ proceeds as we did in the subsonic-supersonic model discussed in the previous section.
The resulting $S$ matrix elements are the following

\bea
    S_{ur,vr} &=& \frac{-\sqrt{3} + \I}{8} (\frac{\hbar \w}{m c_r^2})^{2/3} + O({\w})\ , \\
    S_{vl,vr} &=& \frac{c_l + c_r}{2 \sqrt{c_l c_r}} + \frac{1 - \I\sqrt{3}}{16} \frac{c_l- c_r}{\sqrt{c_l c_r}} (\frac{\hbar\w}{m c_r^2})^{2/3} + O(w)\ , \\
    S_{ul,vr} &=& \frac{c_l - c_r}{2 \sqrt{c_l c_r}} + \frac{1 - \I\sqrt{3}}{16} \frac{c_l + c_r}{\sqrt{c_l c_r}} (\frac{\hbar\w}{m c_r^2})^{2/3} + O(\w)\ , \\
    S_{ur,3l} &=& \frac{\sqrt{3} + \I}{2\sqrt{2}}(c_r^2 - c_l^2)^{1/4} (\frac{m}{\hbar \w c_t})^{1/6} + O(\w^{1/6})\ ,\\
    S_{vl,3l} &=& \frac{1 + \I\sqrt{3}}{4\sqrt{2}} \frac{(c_r - c_l)(c_r^2 - c_l^2)^{1/4}}{\sqrt{c_l c_r}} (\frac{m}{\hbar \w c_r})^{1/6} + O(\w^{1/6})\ , \\
    S_{ul,3l} &=& -\frac{1 + \I\sqrt{3}}{4\sqrt{2}} \frac{(c_r + c_l)(c_r^2 - c_l^2)^{1/4}}{\sqrt{c_lc_r}} (\frac{m}{\hbar \w c_R})^{1/6} + O(\w^{1/6})\ , \\
    \label{m} S_{ur,4l} &=& -\frac{\sqrt{3} + \I}{2\sqrt{2}}(c_r^2 - c_l^2)^{1/4} (\frac{m}{\hbar \w c_r})^{1/6} + O(\w^{1/6})\ ,\\
\label{mm}   S_{vl,4l} &=& -\frac{1 + \I\sqrt{3}}{4\sqrt{2}} \frac{(c_r - c_l)(c_r^2 - c_l^2)^{1/4}}{\sqrt{c_l c_r}} (\frac{m}{\hbar \w c_r})^{1/6} + O(\w^{1/6})\ , \\
\label{mmm}    S_{ul,4l} &=& \frac{1 + \I\sqrt{3}}{4\sqrt{2}} \frac{(c_l + c_r)(c_r^2 - c_l^2)^{1/4}}{\sqrt{c_l c_r}} (\frac{m}{\hbar \w c_r})^{1/6} + O(\w^{1/6})\ .
\eea

Because the $ul$ and the $4l$ mode of the supersonic region have negative norm,  in this case too there will be a nontrivial Bogoliubov transformation mixing annihilation and creation operators and hence particles creation out of the vacuum in the three channels $ur, ul, vl$. The corresponding numbers of particles created at fixed (small) $\omega$ are:
\bea
    n_{ur} &\sim&  \frac{\sqrt{c_r^2 - c_l^2}}{2} (\frac{m}{\hbar \w c_r})^{1/3}\ , \nonumber \\
    n_{ul} &\sim&  \frac{(c_l+c_r)^2}{4c_lc_r}\frac{\sqrt{c_r^2 - c_l^2}}{2} (\frac{m}{\hbar \w c_r})^{1/3}\ , \nonumber  \\
    n_{vl} &\sim&  \frac{(c_r-c_l)^2}{4c_lc_r}\frac{\sqrt{c_r^2 - c_l^2}}{2} (\frac{m}{\hbar \w c_r})^{1/3}\ .    \label{eqn:ch5:particle-production-sonic-supersonic}
\eea

One immediately notices a striking difference with respect to the subsonic-supersonic model. The relevant $S$ matrix elements behave now for small $\omega$ as $\w^{-\frac{1}{6}}$ and therefore the numbers of created particles scale as $\w^{-\frac{1}{3}}$ and not as $\w^{-1}$. The particles production is no longer thermal.

\begin{figure}[h]
    \centering
    \includegraphics[width = 0.5\textwidth]{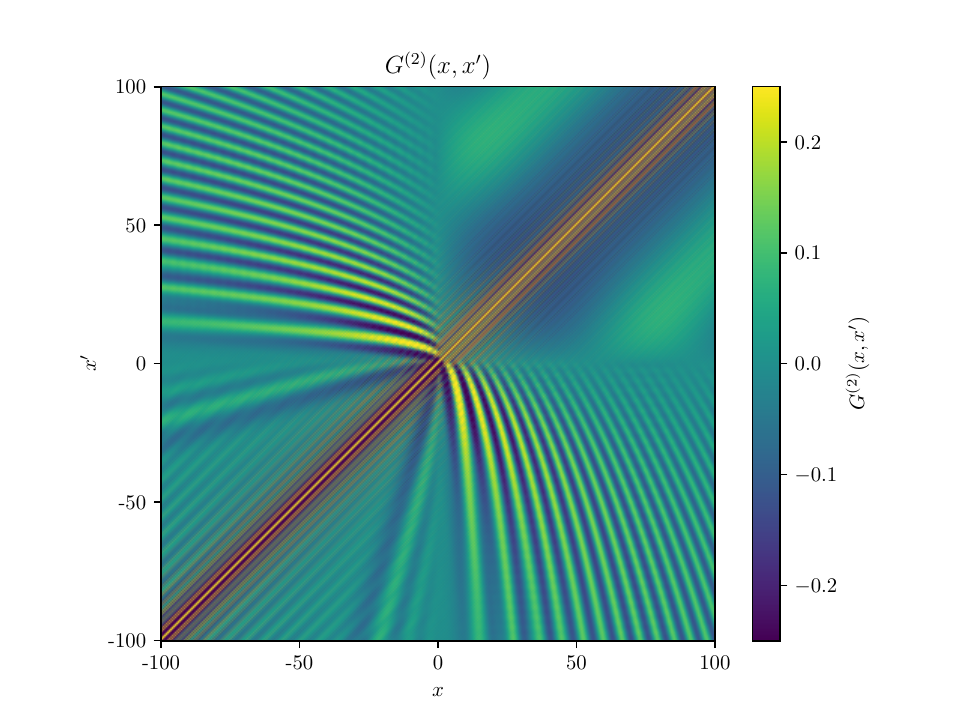}
    \caption{Two-point correlation function for the sonic-supersonic two region model.  We chose the values $c_l = \frac{1}{2}, c_r = 1 = -v, m = 1, n = 1/(4\pi), \hbar=1$.}
    \label{fig:ch5:g2-sonic-supersonic}
\end{figure}

Looking at the corresponding density-density correlation function $G^{(2)}(x,x’)$, Fig. (\ref{fig:ch5:g2-sonic-supersonic}), we see that the characteristic peaks present in  the subsonic-supersonic model (see Fig. (\ref{fig:ch4:twopoint-sub-super})) when the points are taken on opposite sides with respect to the horizon $x=0$ are no longer visible and a sort of ``modulation'' appears in the signal. 

To understand this feature we attempt an analysis similar to what was done at the end of the previous section for the subsonic-supersonic model. Considering the HP correlator (\ref{hpr}) with now $ur$ belonging to the sonic region we clearly see that at low frequency (see Subsection (\ref{ma})) the exponent $k_{ur}x-k_{ul}x'\sim \frac{2}{\xi_r}(\frac{\xi_r\omega}{c_r})^{1/3}x$, i.e. the dependence on the point $x'$ is higher order. The same can be said for the Hawking quanta spectator correlator (\ref{hsr}), where again $k_{ur}x-k_{vl}x'\sim \frac{2}{\xi_r}(\frac{\xi_r\omega}{c_r})^{1/3}x$. Based on these considerations and using the leading order low energy expansions in Subsection (\ref{ma}), Appendix (\ref{appendixA}) and the matrix elements (\ref{m}, \ref{mm}, \ref{mmm}) an approximate form for the correlator when one point is outside ($x$) and the other inside ($x'$) is 
\bea \label{g2} G^{(2)}_{HP+HS} &\sim&A\Big[ -3\sqrt{3}\left( -y_{max}^2\frac{\sin y_{max}x}{x}-2y_{max}\frac{\cos y_{max}x}{x^2}+2\frac{\sin y_{max}x}{x^3}\right)\nonumber  \\ &+& 3\left( -y_{max}^2\frac{\cos y_{max}x}{x}+2y_{max}\frac{\sin y_{max}x}{x^2}+2\frac{\cos y_{max}x-1}{x^3}\right)\Big] \ ,
\eea
where $A=-\frac{\sqrt{c_lc_r}}{8\sqrt{3}\sqrt{c_r^2-c_l^2}} \xi_l^{1/2}\xi_r^{3/2}$ and $y_{max}=\frac{2}{\xi_r}(\frac{\xi_r\omega_{max}}{c_r})^{1/3}$. We see that in this approximation the signal presents (damped) oscillations only in $x$, a feature that is in qualitative agreement with the (almost) vertical modulation present in the bottom right quadrant of Fig. (\ref{fig:ch5:g2-sonic-supersonic}).

On the other hand the feature in the interior region of Figs. (\ref{fig:ch4:twopoint-sub-super})   and (\ref{fig:ch5:g2-sonic-supersonic}) (i.e. $x,x’$ both $<0$), describing correlations between partners and spectators, is qualitatively similar to the one found in the subsonic-supersonic model showing a distinct negative peak.

\section{Three regions model: subsonic-sonic-supersonic}
\label{trm}

\begin{figure}[h]
\includegraphics[width=0.5\columnwidth]{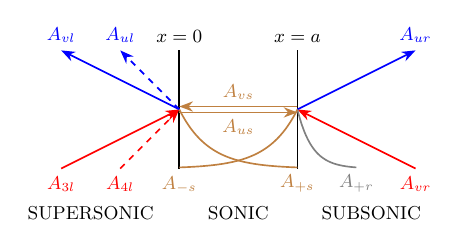} 
\caption{Matching between incoming modes (in red) and outgoing ones (in blue).
}
\label{fquattro}
\end{figure}

The last model we discuss consists of three regions: a subsonic one (i.e $|v|<c_r$) for $x>a$, a sonic one (i.e. $|v|=c_s$) for $0<x<a$ and a supersonic one (i.e. $|v|>c_l$) for $x<0$ with $c_r>c_s>c_l$. The model contains therefore an extended sonic region. In a gravitational language we have a thick horizon of width $a$.
The modes involved are depicted in Fig. (\ref{fquattro}).

In red we have the ingoing modes with momenta $k_{3l}, k_{4l}, k_{vr}$, in blue the outgoing ones with momenta $k_{vl}, k_{ul}, k_{ur}$. Note that in the intermediate sonic region, besides the two propagating modes with momenta $k_{us}$ and $k_{vs}$, there is not only the decaying mode $k_{+s}$ but also the growing one $k_{-s}$. This because of the finite extension of this region.

\begin{figure}[h]
\includegraphics[width=0.5\columnwidth]{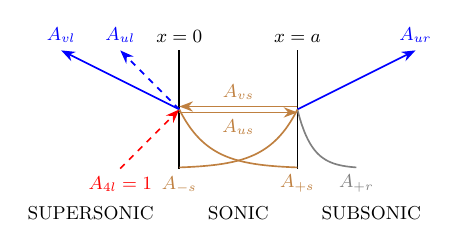} 
\caption{Relevant amplitudes for the $\phi^{in}_{4l}$ in mode.}
\label{f4l}
\end{figure}

\begin{figure}[h]
\includegraphics[width=0.45\columnwidth]{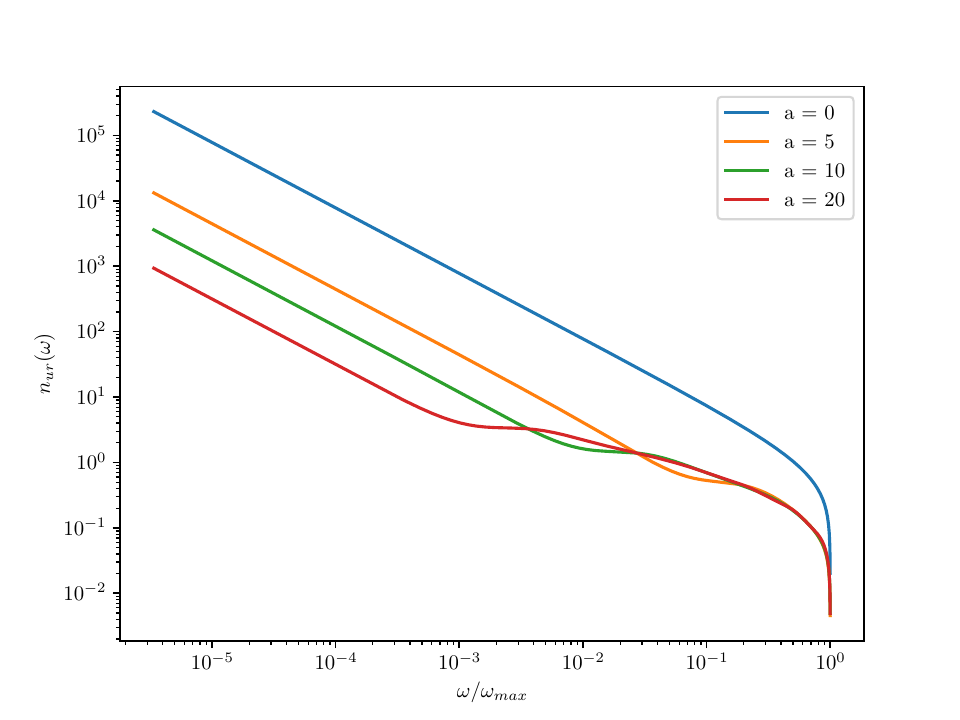} 
\caption{(Log) Plot of $n_{ur}(a)$, number of Hawking particles emitted in the subsonic region  for $a=0,5,10,20$. Here and in the figures that follow we chose the values $c_r = 2, c_s = 1 = -v, c_l = 0.5, m = 1, n = 1/(4\pi), \hbar=1$.}
\label{nur3}
\end{figure}

\begin{figure}[h]
\includegraphics[width=0.45\columnwidth]{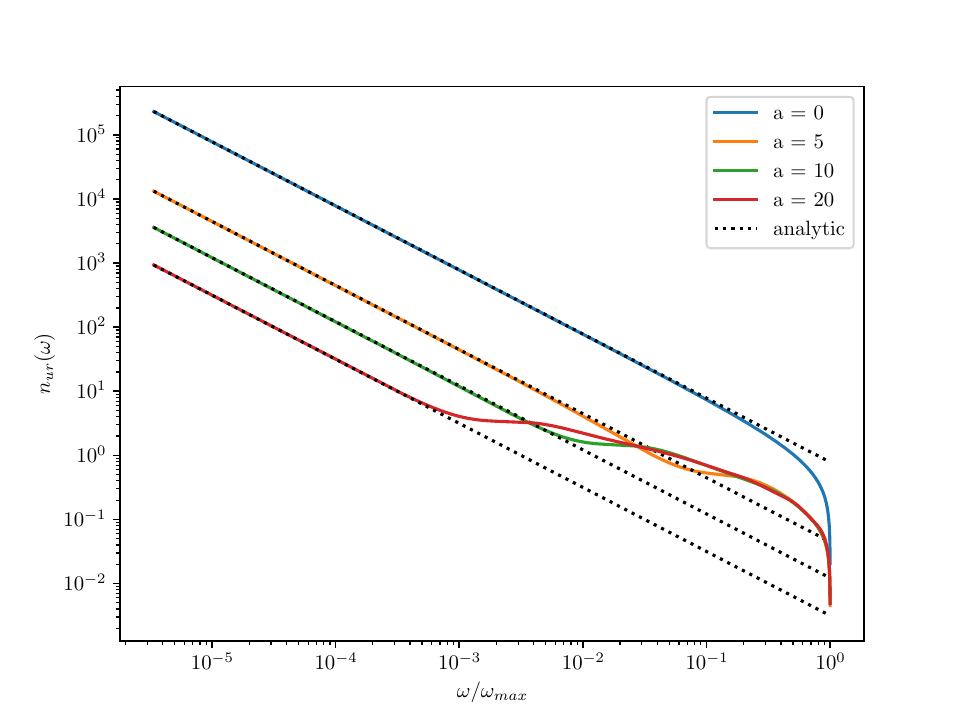} 
\caption{Comparison between the numerical (continuous lines) and the analytic (leading in $\omega$) result (dotted lines) in (\ref{han}) for  $n_{ur}(a)$.}
\label{nur3na}
\end{figure}

In this model we have to impose two sets of matching conditions on the modes and their derivatives, one at $x=0$ between the supersonic and the sonic regions and the other at $x=a$ between the sonic and the subsonic one.
In analogy with eq. (\ref{qcinque}), we have that the matching at $x=0$ imposes the following relations between the supersonic and the sonic amplitudes
\be \label{pm} M_lA_l=M_sA_s\ , \ee
where $M_l$ is given by
\be
\label{eq:wl}
M_{l}=\left(
     \begin{array}{cccc}
       D_{vl} & D_{ul} & D_{3l} & D_{4l}\\
       ik_{vl}D_{vl} & ik_{ul}D_{ul} & ik_{3l}D_{3l} & ik_{4l}D_{4l} \\
       E_{vl} & E_{ul} & E_{3l} & E_{4l} \\
       ik_{vl}E_{vl} & ik_{ul}E_{ul} & ik_{3l}E_{3l} & ik_{4l}E_{4l}  \\
\end{array}\right)\,. 
\end{equation}
$M_s$ has a similar form with $vl, ul, 3l, 4l$ replaced by $vs, us, +s , -s$. On the other hand, the matching at $x=a$ between the sonic and subsonic regions requires that
\be\label{sm} N_sA_s=N_rA_r\ ,\ee
where $N_S$ is given by
\be \label{ns}
N_{s}=\left(
     \begin{array}{cccc}
       D_{vs}e^{ik_{vs}a} & D_{us}e^{ik_{us}a} & D_{+s}e^{ik_{+s}a} & D_{-s}e^{ik_{-s}a}\\
       ik_{vs}D_{vs}e^{ik_{vs}a} & ik_{us}D_{us}e^{ik_{us}a} & ik_{+s}D_{+s}e^{ik_{+s}a} & ik_{-s}D_{-s}e^{ik_{-s}a} \\
       E_{vs}e^{ik_{vs}a}& E_{us}e^{ik_{us}a} & E_{+s}e^{ik_{+s}a} & E_{-s}e^{ik_{-s}a} \\
       ik_{vs}E_{vs}e^{ik_{vs}a} & ik_{us}E_{us}e^{ik_{us}a} & ik_{+s}E_{+s}e^{ik_{+s}a} & ik_{-s}E_{-s}e^{ik_{-s}a}  \\
\end{array}\right)\, 
\ee
and $N_r$ is obtained by replacing $us, vs, +s$ and $-s$ by $ur, vr, +r$ and $-r$ and setting $A_{-r} =0$.
Combining eqs. (\ref{pm}), (\ref{sm}) we can eliminate the four sonic amplitudes and get
\be\label{me3}  M_lA_l=M_sN_{s}^{-1}N_r A_r\ .\ee   
A brief description of the numerical analysis followed to solve  the above equations is given in Appendix (\ref{appendixC}). The construction of the `in' and `out' bases proceeds in the usual way. 
For example in Fig. (\ref{f4l})
is depicted the construction of the $\phi^{in}_{4l}$ in mode.
The associated `in' and `out' creation and annihilation operators are connected by the $S$ matrix as in Eq. (\ref{Ssupsup})  and the number of created particles follows Eq.  (\ref{sund}).
In Fig. (\ref{nur3})
is depicted the number $n_{ur}(a)$ of the Hawking particles emitted in the subsonic region for various values of the width $a$ of the sonic region. 
One immediately notices that this number decreases as $a$ increases. An analytical estimate of $n_{ur}$ at small $\omega$
can be given as (see Appendix (\ref{appendixD})  for details of the derivation)
\bea \label{han} n_{ur}(a) &=& |A_{ur}(a)|^2 \ ,   \\
 A_{ur} (a)&=& 
    \sqrt{\frac{2c_l \xi_l c_r (c_r-c_s)\left(c_s^2-c_l^2\right)^{3/2}}{\omega(c_r+c_s)}}
    \frac{c_l \xi_l \left( \sqrt{c_r^2-c_s^2}-i\sqrt{c_s^2-c_l^2} \right)-i S_a}{\ S_a \left(a \sqrt{c_r^2-c_s^2} +c_l \xi_l \right) \sqrt{c_s^2-c_l^2}+c_l^2 \xi_l^2 \left(c_r^2-c_l^2\right)},\ \ \ \  \nonumber
\eea
where the sonic damping factor $S(a)$ is defined as
\be \label{df}
S(a) \equiv 2 a  \sqrt{c_r^2-c_s^2} \sqrt{c_s^2-c_l^2}\ . \ee
For $a=0$, $S(a)=0$ and Eq. (\ref{han}) reduces to Eqs. (\ref{uuuu}), (\ref{uuu})  of the subsonic-supersonic model as expected. From Eq. (\ref{han}) we see that $n_{ur}(a)$ scales as $\frac{1}{\omega}$ like we had in the subsonic-supersonic model. The accuracy of this approximation can be appreciated in Fig. (\ref{nur3na})
where the analytical
result is represented by the dotted lines while the continuous lines correspond to the numerical one. 

\begin{figure}[h]
\includegraphics[width=0.45\columnwidth]{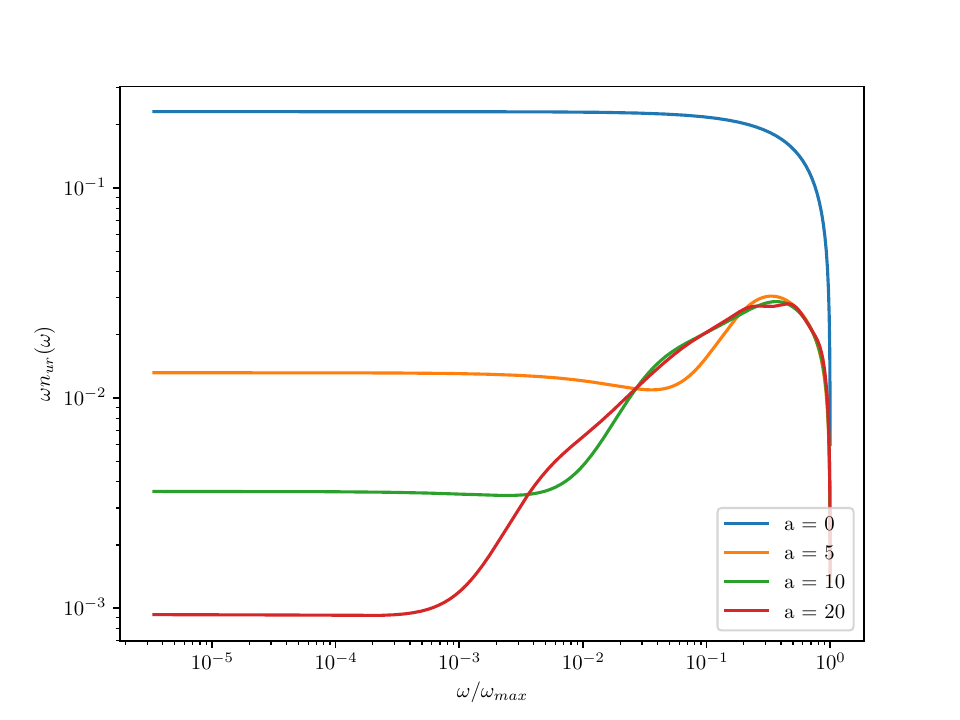} 
\caption{Plot of $\w n_{ur}$ for $a=0,5,10,20$. We see the plateau (corresponding to $n_{ur}\sim \frac{1}{\omega}$), which decreases as $a$ increases, and then a transition to a $n_{ur}\sim \frac{1}{\omega^{1/3}}$ behaviour (see also Ref. \cite{pdbf} and Fig. (\ref{nur313na})).}
\label{nur31na}
\end{figure}

In Fig. (\ref{nur31na})
we plot $\w n_{ur}$; the plateau corresponds to the expected $\frac{1}{\omega}$ behaviour at small $\omega$, but then a transition to a $\frac{1}{\omega^{1/3}}$ regime occurs which is typical of a sonic region (see Eqs.  (\ref{eqn:ch5:particle-production-sonic-supersonic})).

\begin{figure}[h]
\includegraphics[width=0.45\columnwidth]{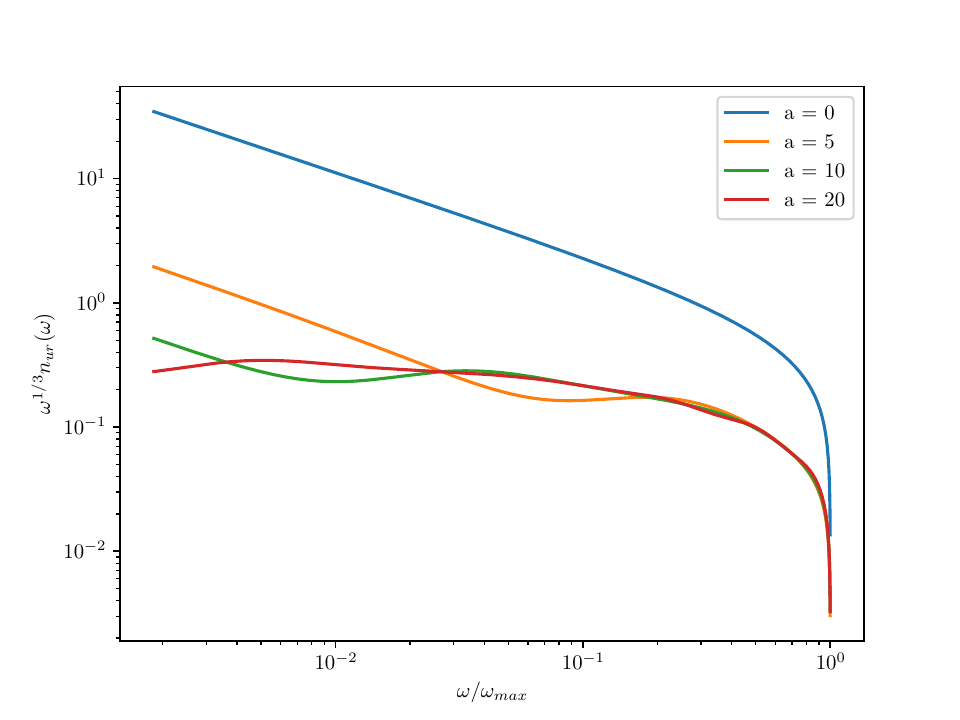} 
\caption{Plot of $\omega^{1/3}n_{ur}$ for $a=0,5,10,20$. 
}
\label{nur313na}
\end{figure}

This figure should be compared with the analogous one, Fig. (\ref{planck0-nur-ss}),
of the subsonic-supersonic model where the $\frac{1}{\omega}$ behaviour lasts till the emission drops out as $\omega$ approaches $\omega_{max}$. In Fig. (\ref{nur313na})
we plot $\omega^{1/3}n_{ur}$, showing that the $\frac{1}{\omega^{1/3}}$ region increases (and correspondingly the $\frac{1}{\omega}$ one decreases) as the width a increases. \footnote{In the limit $a\to +\infty$ we recover the two regions supersonic-sonic model: the $\frac{1}{\omega}$ term in the particles number completely disappears, consistently with the results in subsection (\ref{ss}).}

\begin{figure}[h]
    \includegraphics[width=0.45\textwidth]{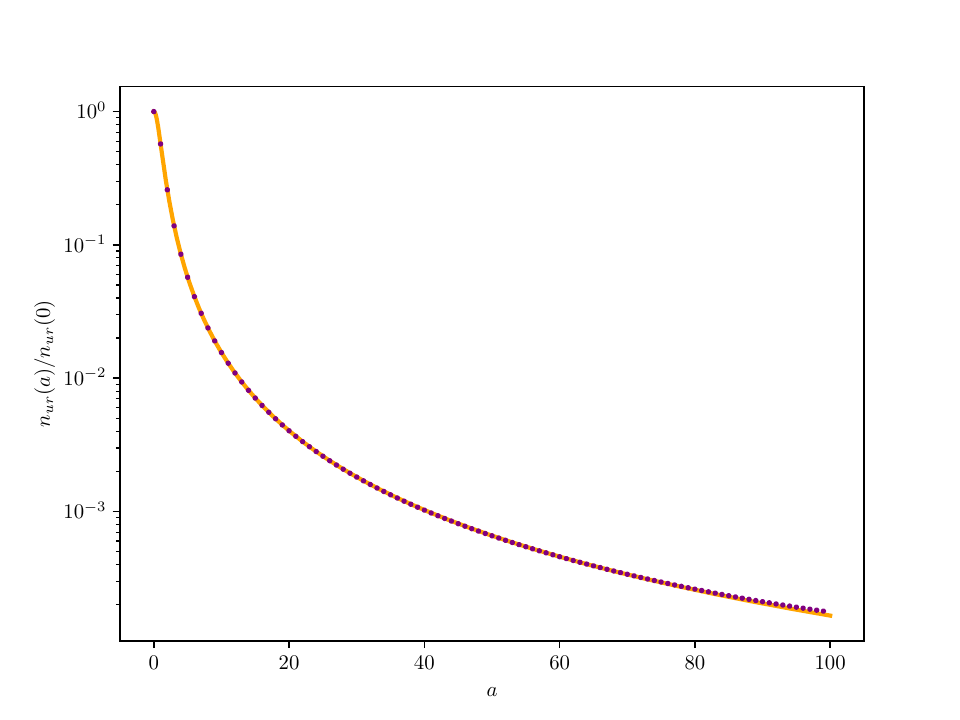}
    \caption{Numerical (red dotted line) and analytical plot (continuous line) of the (low-frequency) gray-body factor $\Gamma$. 
    }
    \label{fig:lettera-T/T0-act}
\end{figure}

\begin{figure}[h]
    \begin{subfigure}{0.8\textwidth}
    \includegraphics[width=0.535\columnwidth]{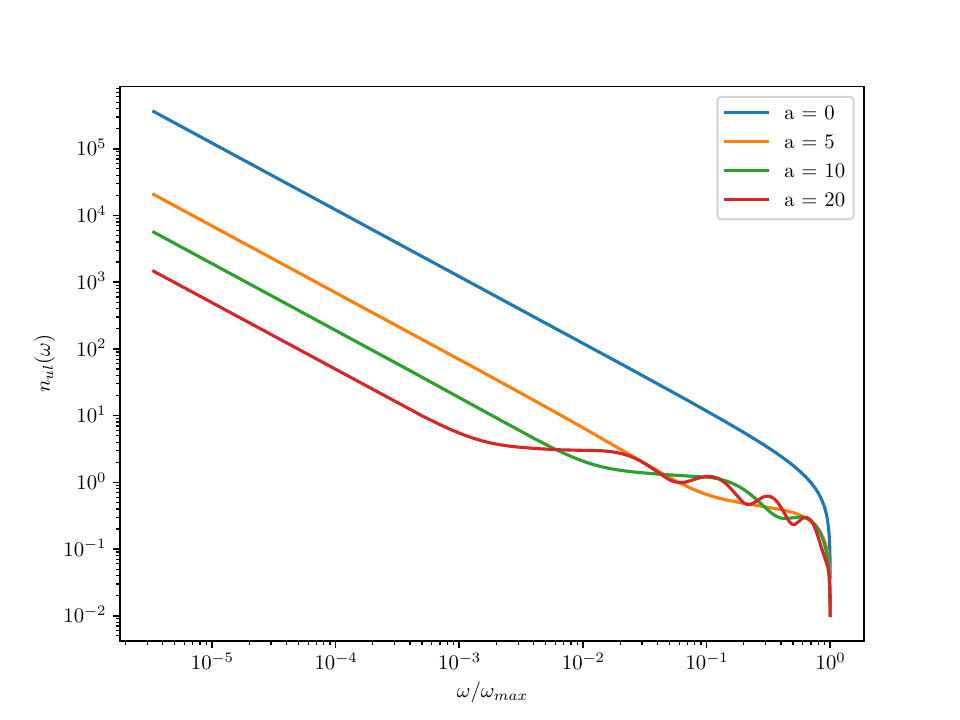} 
\caption{}
\end{subfigure}
    \begin{subfigure}{0.8\textwidth}
    \includegraphics[width=0.535\columnwidth]{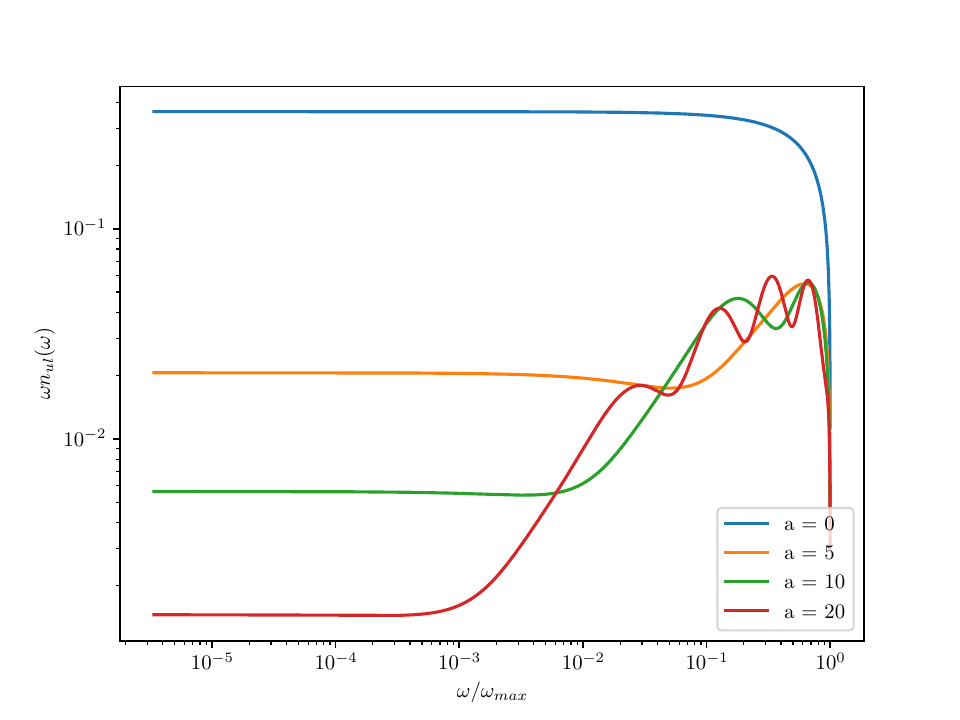} 
\caption{ 
}
\end{subfigure}
    \begin{subfigure}{0.8\textwidth}
    \includegraphics[width=0.535\columnwidth]{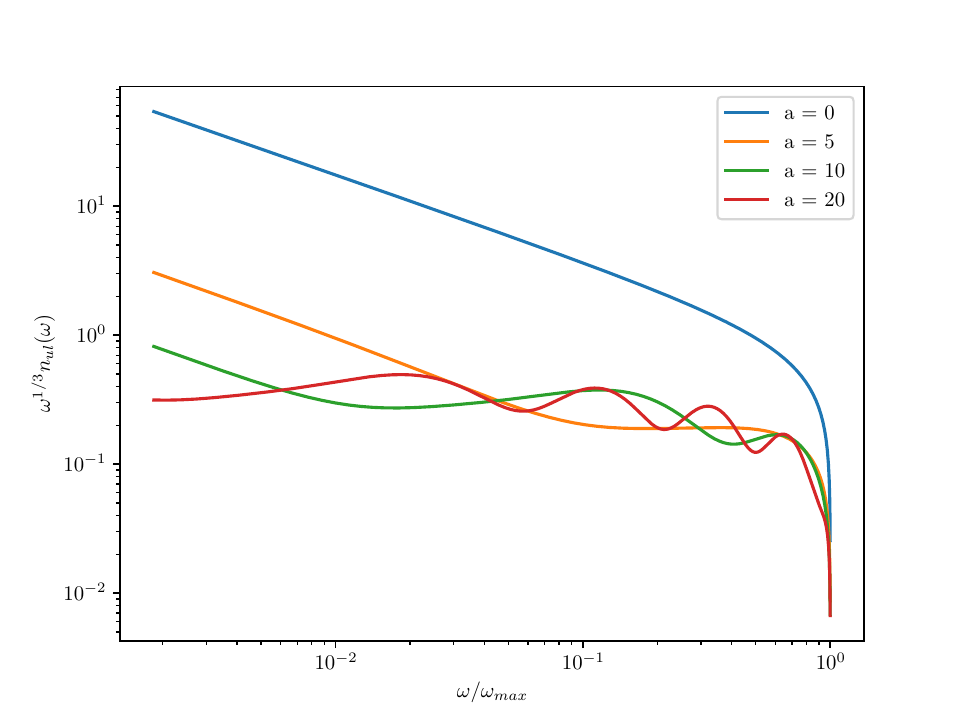} 
\caption{
}
\end{subfigure}
\caption{ (Log) Plots of a) $n_{ul}(a)$, b) $\w n_{ul}$, 
c) $\omega^{1/3}n_{ul}$ for $a=0,5,10,20$. }
\label{nul3-nul31na-nul313na}
\end{figure}

\begin{figure}[h]
\begin{subfigure}{0.8\textwidth}
\includegraphics[width=0.535\columnwidth]{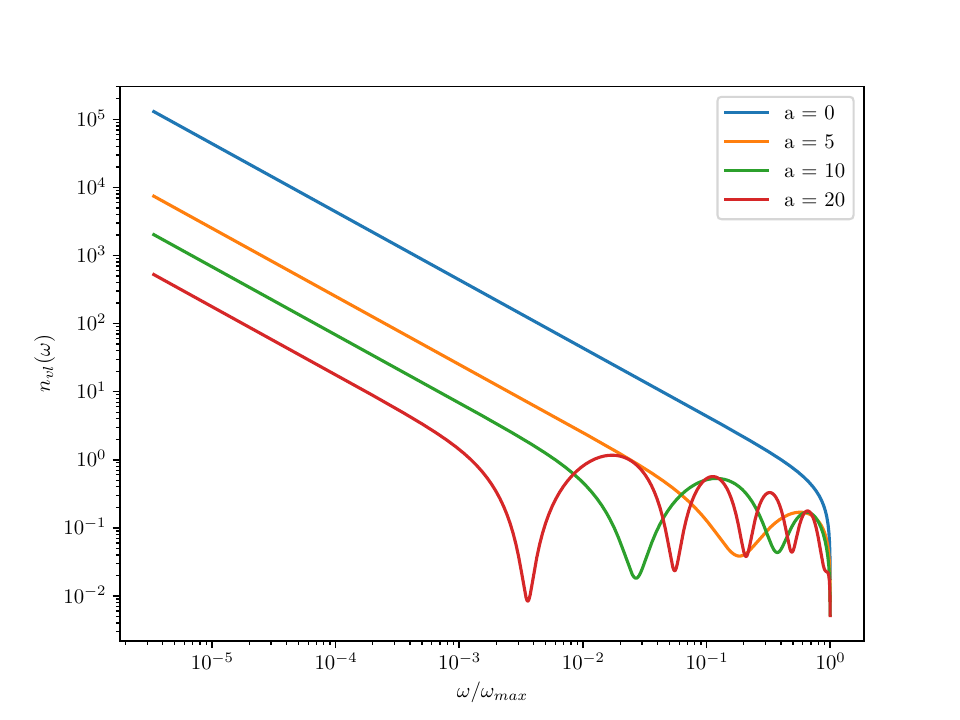} 
\caption{}
\end{subfigure}
\begin{subfigure}{0.8\textwidth}
\includegraphics[width=0.535\columnwidth]{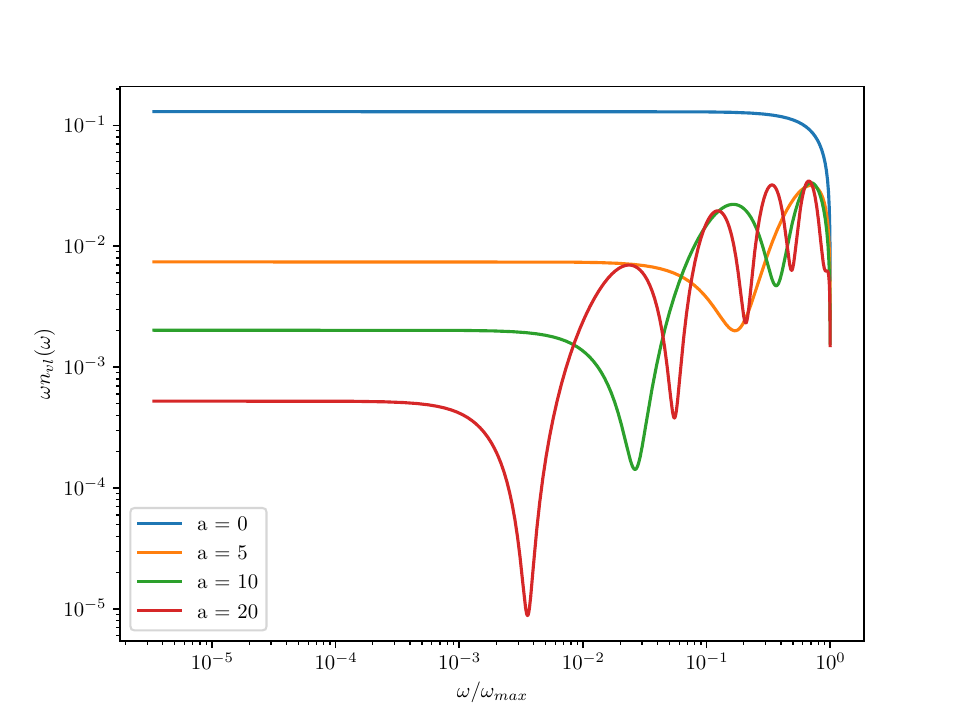} 
\caption{ 
}
\end{subfigure}
\begin{subfigure}{0.8\textwidth}
\includegraphics[width=0.535\columnwidth]{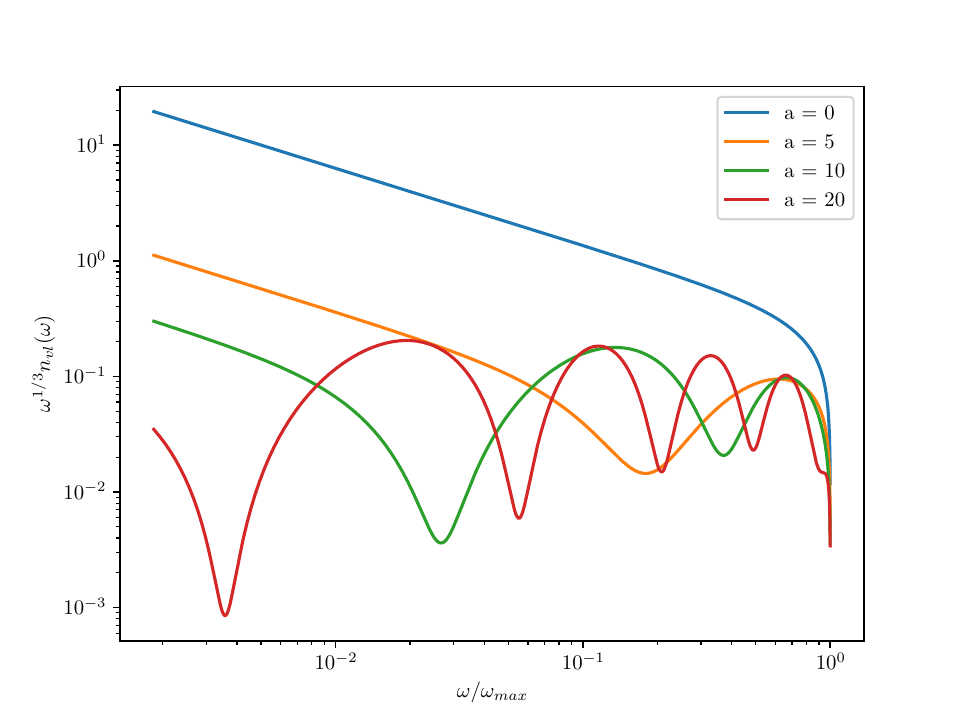} 
\caption{
}
\end{subfigure}
\caption{(Log) Plots of a) $n_{vl}(a)$, b) $\w n_{vl}$, c) $\omega^{1/3}n_{vl}$ 
for $a=0,5,10,20$. }
\label{nvl3-nvl31na-nvl313na}
\end{figure}

To the $\frac{1}{\omega}$ phase one can associate a thermal emission as in Eq. (\ref{td}) at an effective temperature $T_H$ of Eq. (\ref{htt}) damped by a low frequency gray-body factor $\Gamma(\omega,a)$
\be \label{httr}
   n_{ur} (\omega,a) \sim \frac{\Gamma k_B T_H}{\hbar\omega}\ , \ee
where $\Gamma= \frac{n_{ur}(a)}{n_{ur}(a=0)}$, due to the backscattering of the modes occurring inside the sonic region. From Fig. (\ref{nur31na}) we infer that at small frequency $\Gamma$ is independent of $\omega$, consistently with what was found for acoustic BHs in \cite{abfp, afb}, and it scales as $\frac{1}{a^2}$ for $a\gg 1$ (this can be seen from Eq. (\ref{han})). The behaviour of $\Gamma$ as a function of $a$ is depicted in Fig. (\ref{fig:lettera-T/T0-act}), where the solid line corresponds to the analytical result obtained using Eq. (\ref{han}), while the dotted line corresponds to the numerical results.
In Fig. (\ref{nur313na})  another significant difference appears: the $\frac{1}{\omega^{1/3}}$ phase is characterised by oscillations. These features are much more enhanced when one performs a similar analysis of the signals corresponding to the particle production in the other two outgoing channels $n_{ul}$ (partners) and $n_{vl}$ (spectators) that are present in the supersonic region. See Figs. (\ref{nul3-nul31na-nul313na}), 
(\ref{nvl3-nvl31na-nvl313na}).  


\begin{figure}[h]
\includegraphics[width=0.45\columnwidth]{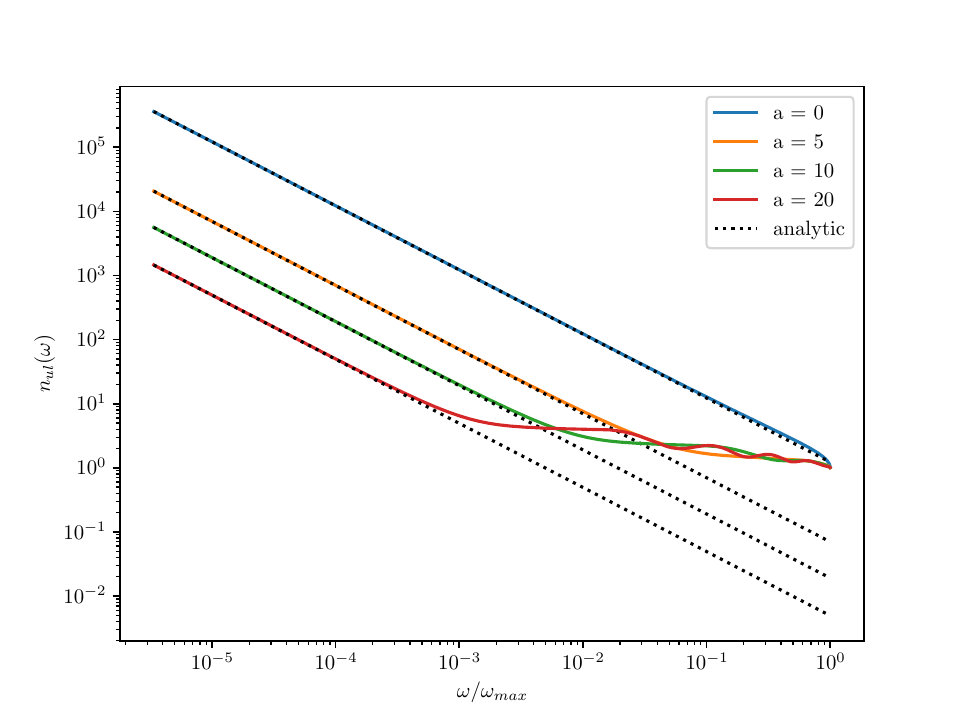} 
\caption{Comparison between the numerical (continuous lines) and the (eading in $\omega$ analytical  result (dotted lines) in (\ref{hanu}) for  $n_{ul}(a)$.}
\label{nul3na}
\end{figure}

\begin{figure}[h]
\includegraphics[width=0.45\columnwidth]{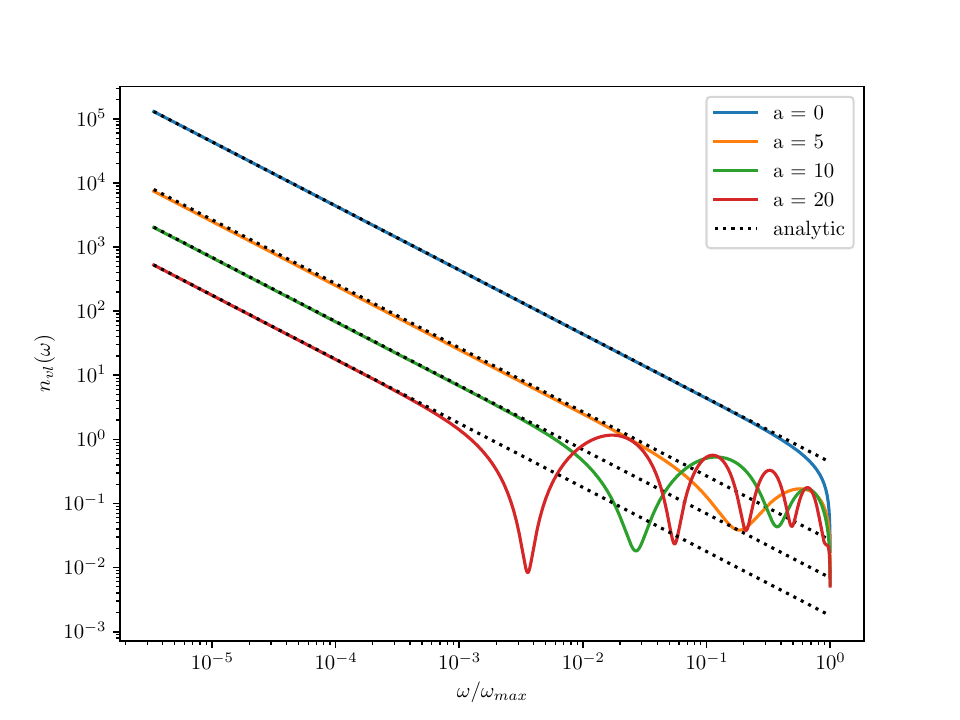} 
\caption{Comparison between the numerical (continuous lines) and the leading in $\omega$ analytical result (dotted lines) in (\ref{hanv}) for  $n_{vl}(a)$.}
\label{nvl3na}
\end{figure}

An analytical estimate of $n_{ul}, n_{vl}$ at small $\omega$
gives (see Appendix (\ref{appendixD}))
\bea \label{hanu} &&n_{ul}(a) \sim |A_{ul}(a)|^2 \ ,   \\ \nonumber
&&\begin{split}
    A_{ul}(a) =
    &\sqrt{\frac{\xi_l}{2 \omega} }
    \frac{1}{\sqrt{c_r^2-c_s^2}
    \left[
    S_a \sqrt{c_s^2-c_l^2} \left(a \,  \sqrt{c_r^2-c_s^2}  
    +c_l \xi_l\right)
    + c_l^2 \xi_l^2 \left(c_r^2-c_l^2 \right) 
    \right]^2
    }
    \\
    &\times
    \left[
    -(c_r +c_l)(c_r-c_s)(c_s^2-c_l^2)^{3/4}
    \left(c_l^3 \xi_l^3 (c_r^2-c_l^2) \left(\sqrt{c_r^2-c_s^2} -i \sqrt{c_s^2-c_l^2} \right)
    \right.
    \right.
    \\
    &-a \sqrt{c_s^2-c_l^2}
    \left\{
    S_a \sqrt{c_r^2-c_s^2}
    \left[
    iS_a -c_l \xi_l 
    \left(
    \sqrt{c_r^2-c_s^2}-3 i \sqrt{c_s^2-c_l^2} 
    \right)
    \right]
    \right.
    \\
    &\left.
    \left.
    \left.
    +2 i c_l^2 \xi_l^2
    \left[
    c_r^2 \left( \sqrt{c_r^2-c_s^2} +i \sqrt{c_s^2-c_l^2} \right)
    +c_s^2 \left( \sqrt{c_r^2-c_s^2} -i \sqrt{c_s^2-c_l^2} \right)
    -2 c_l^2 \sqrt{c_r^2-c_s^2}
    \right]\right\}\right)\right],
\end{split}
\eea
\bea \label{hanv} && n_{vl}(a) = |A_{vl}(a)|^2 \ ,   \\
 \nonumber 
 && 
 \begin{split}
    A_{vl}(a) =
    &\sqrt{\frac{\xi_l}{2 \omega} }
    \frac{1}{\sqrt{c_r^2-c_s^2}
    \left[
    S_a \sqrt{c_s^2-c_l^2} \left(a \,  \sqrt{c_r^2-c_s^2}  
    +c_l \xi_l\right)
    + c_l^2 \xi_l^2 \left(c_r^2-c_l^2 \right) 
    \right]^2
    }
    \\
    &\times
    \left[
    (c_r-c_l)(c_r-c_s)(c_s^2-c_l^2)^{3/4}
    \left(c_l^3 \xi_l^3 (c_r^2-c_l^2) \left(\sqrt{c_r^2-c_s^2} -i \sqrt{c_s^2-c_l^2} \right)
    \right.
    \right.
    \\
    &-a \sqrt{c_s^2-c_l^2}
    \left\{
    S_a \sqrt{c_r^2-c_s^2}
    \left[
    iS_a -c_l \xi_l 
    \left(
    \sqrt{c_r^2-c_s^2}-3 i \sqrt{c_s^2-c_l^2} 
    \right)
    \right]
    \right.
    \\
    &\left.
    \left.
    \left.
    +2 i c_l^2 \xi_l^2
    \left[
    c_r^2 \left( \sqrt{c_r^2-c_s^2} +i \sqrt{c_s^2-c_l^2} \right)
    +c_s^2 \left( \sqrt{c_r^2-c_s^2} -i \sqrt{c_s^2-c_l^2} \right)
    -2 c_l^2 \sqrt{c_r^2-c_s^2}
    \right]\right\}\right)\right],
\end{split} 
\eea
where  the damping factor $S(a)$ was defined in Eq. (\ref{df}).
For $a=0$ Eqs. (\ref{hanu}), (\ref{hanv}) reduce to Eqs. (\ref{uuuu}), (\ref{ulul}), (\ref{vlvl})  of the subsonic-supersonic model as expected. 
The accuracy of this approximation can be appreciated in Figs. (\ref{nul3na}), (\ref{nvl3na}).

\begin{figure}[h]
    \begin{center}
    \begin{subfigure}{0.45\textwidth}
        \includegraphics[width=1.1\linewidth]{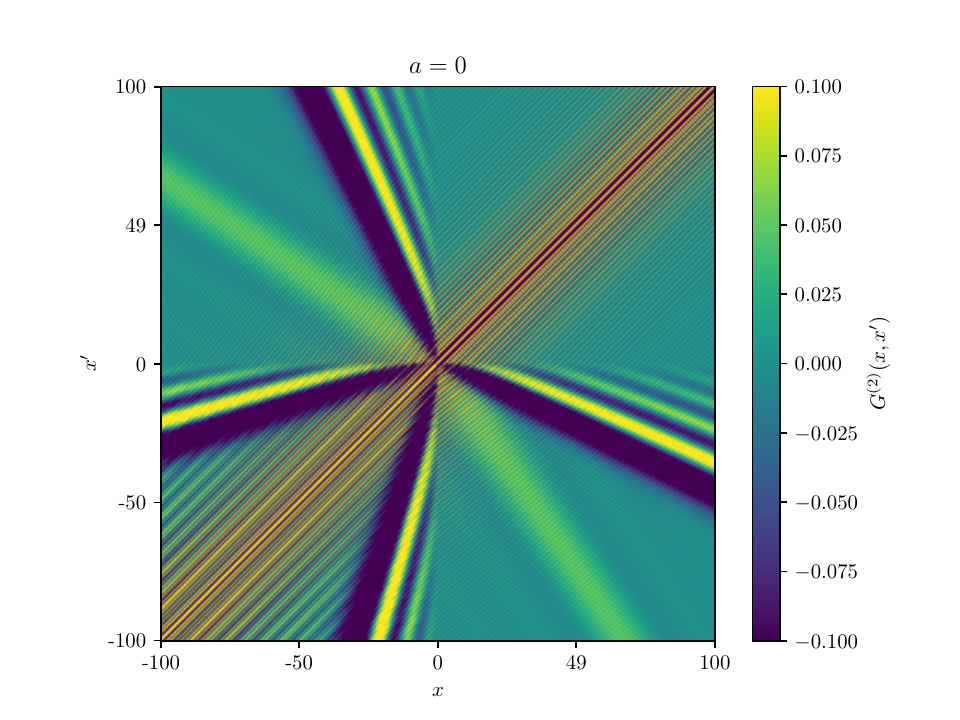}
        \caption{$a = 0$}
    \end{subfigure}
    \hfill
    \begin{subfigure}{0.45\textwidth}
        \includegraphics[width=1.1\linewidth]{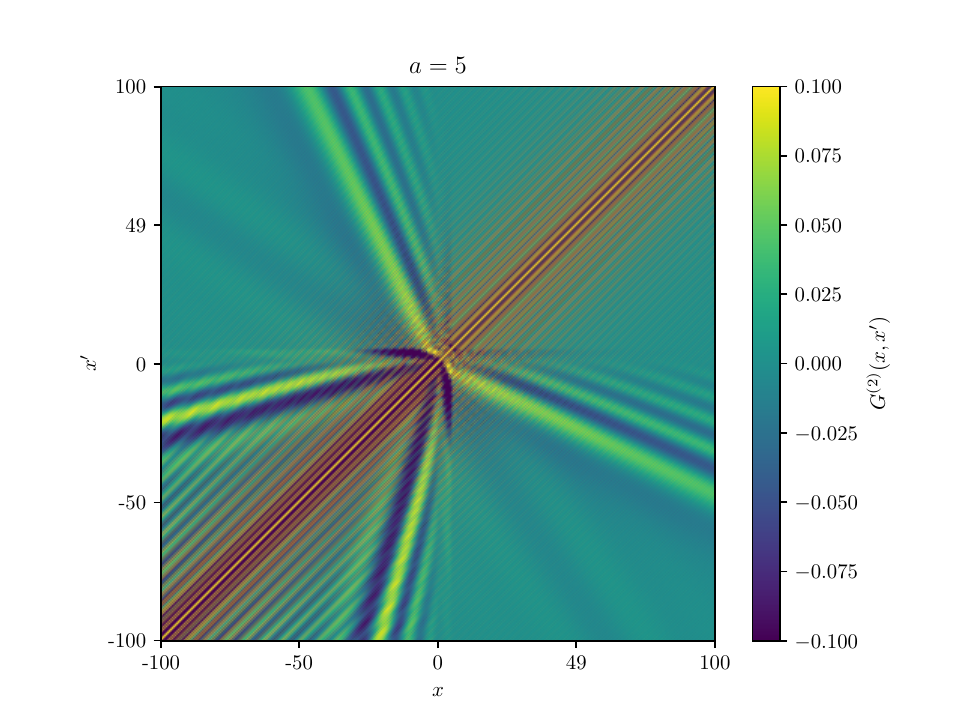}
        \caption{$a = 5$}
    \end{subfigure}

    \vspace{0.5em}

    \begin{subfigure}{0.45\textwidth}
        \includegraphics[width=1.1\linewidth]{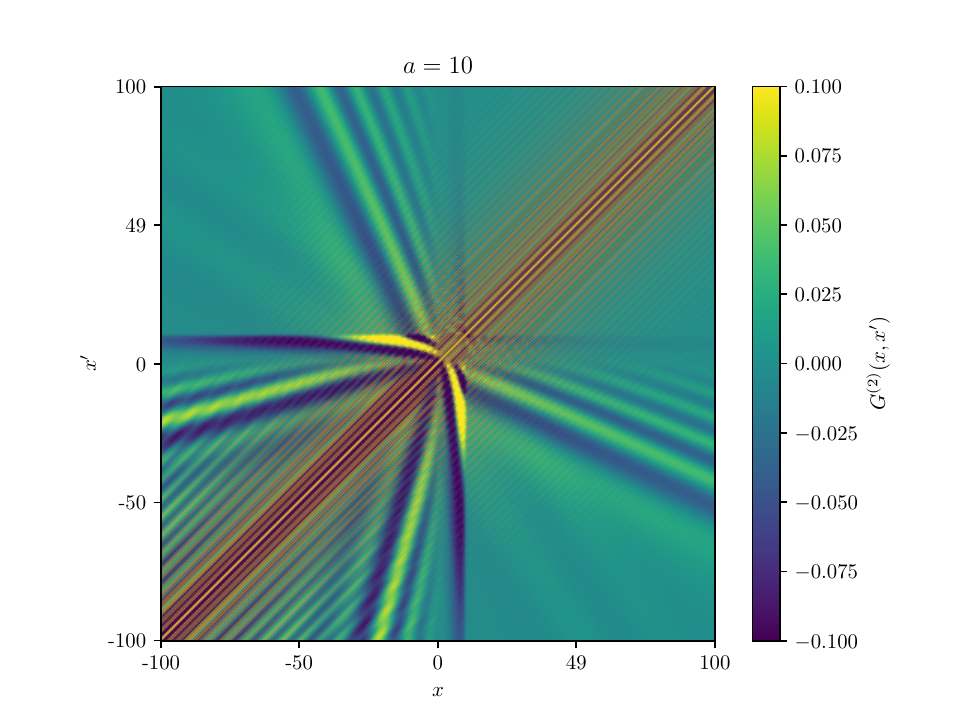}
        \caption{$a = 10$}
    \end{subfigure}
    \hfill
    \begin{subfigure}{0.45\textwidth}
        \includegraphics[width=1.1\linewidth]{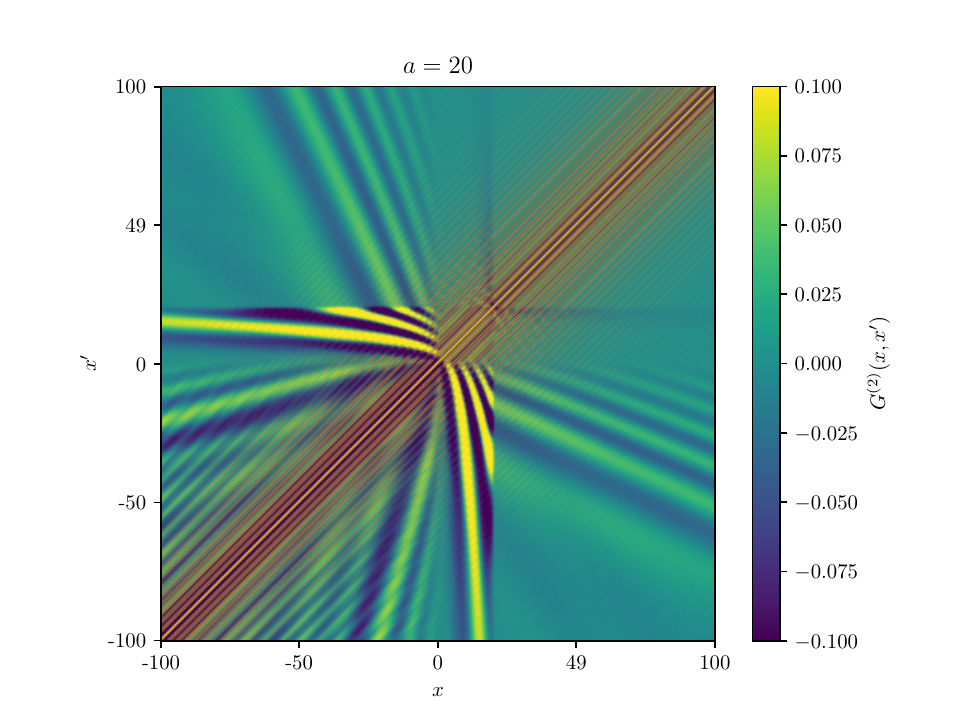}
        \caption{$a = 20$}
    \end{subfigure}
    \end{center}
\caption{Two-point density-density correlation function for the thick horizon model at different values for the sonic region width $a$. We can see how the Hawking-partner correlations decay when $a$ grows and new correlations inside the sonic region quickly emerge. 
    }
    \label{fig:ch6:twopoint-corr-manya}
\end{figure}

Finally, the density-density correlation function $G^{(2)}(x,x')$ for this model for various values of $a$ is given in Fig. (\ref{fig:ch6:twopoint-corr-manya}). One clearly sees the dramatic change occurring in the lower right quadrant (equivalently, in the upper left one) as $a$ increases. The main signal corresponding to the negative Hawking-partner correlation gets weaker and is replaced by a strong almost vertical oscillation between negative and positive values occouring inside the sonic region, reflecting what we saw in the sonic-supersonic model (see Fig. (\ref{fig:ch5:g2-sonic-supersonic})). Outside this region, in the subsonic one, we have a set of parallel oblique fringes: while the slope is the same as for the Hawking quanta-partner correlator of the sub-super ($a=0$) case, as $a$ increases the leading (small $\omega$)  $\frac{1}{\omega}$ contribution to the correlator becomes less important and is replaced by the subleading $\frac{1}{\omega^{1/3}}$ one that, as we have seen (Figs. (\ref{nur3na}, \ref{nur31na}, \ref{nur313na}, \ref{nul3-nul31na-nul313na})) is accompanied by oscillations.


\section{Conclusions}

In this paper we have discussed, using a simple model of an acoustic BH based on a one-dimensional stepwise homogeneous BEC, the effect a finite width of the horizon has on Hawking like radiation in these systems.
The analysis was focussed on the numbers of the created different particles (Hawking quanta, partners, spectators) and on the signature of Hawking radiation in the correlation functions. These latter represent at the moment the basic experimental tool
to detect in laboratory this effect.

We started by analysing waves propagation in a BEC undergoing sonic motion, i.e. the velocity of the flow equals exactly the speed of sound everywhere. Due to the dispersive character of the dispersion relation, it is possible to have waves that counter propagate  with respect to the flow. Note that in the hydrodynamical approximation this behaviour disappears and only co-propagating motion is possible.

In the framework of one-dimensional BEC model, the standard one so far considered in the literature to discuss Hawking radiation consists of two semi-infinite regions, one subsonic the other supersonic, glued along an infinitely thin discontinuity which plays the role of the horizon of the sonic BH. Particles (phonons) creation occurs which is approximately thermal and the density-density correlation shows three characteristic peaks, two for the in-out correlator, the main one corresponding to the Hawking quanta-partners correlations, and one for the in-in correlator (partner-spectator).
\\ \indent
As a warm up exercise, in this model we replaced the subsonic semi-infinite region by a sonic one and showed that the results change dramatically. There is still particles creation, but this is no longer thermal and the above features of the in-out correlator (two distinct peaks) disappear, being replaced by a series of parallel, almost vertical, oscillating fringes.
\\ \indent
We then proceeded to the construction of our thick horizon model of an acoustic BEC BH consisting of two semi-infinite regions, one subsonic and the other supersonic, separated by a sonic region of finite width. We found that the emission in the asymptotic regions is still thermal at low frequency with an amplitude which is reduced compared to the case where the size of the sonic region goes to zero. As the frequency increases the emission is no longer thermal and is characterised by oscillations in the number of the emitted particles. Furthermore, by looking at the in-out correlator, as the width of the sonic region increases, the two characteristic peaks fade out and a series of parallel oblique bands appears.
\\ \indent
From this study one can conclude that infinitely thin horizons are therefore much more efficient to identify Hawking-like radiation in BEC. This not only because the grey-body factor, caused by the backscattering of the modes occurring inside the extended sonic region, reduces the intensity of the signal, but also because the characteristic imprint of Hawking radiation in the correlation functions in the form of peaks which is the experimental smoking gun to reveal the presence of this exotic effect gets washed out.

\textbf{Acknowledgments:} 
A.F., M.D.V. and D.P. acknowledge partial financial support by the Spanish Grant 
PID2023-149560NB-C21 funded by MCIN/AEI/10.13039/501100011033 and FEDER, European Union, and the Severo Ochoa Excellence Grant CEX2023-001292-S. D.P. acknowledges the `Atracci\'o de Talent' program of the University of Valencia for a doctoral grant. M.D.V. acknowledges `Provincia San Domenico in Italia' for a doctoral grant.


\begin{appendix}

\section{$D$ and $E$ coefficients for subosnic, supersonic and sonic flows}
\label{appendixA}

In this Appendix we give the prefactors $D,E$ at small frequencies for all modes for subsonic
\bea
    \label{eqn:ch4:subsonic-D-start}
    D_v &=& \frac{1}{\sqrt{\xi \w}} + \frac{1}{2}\frac{\sqrt{\xi \w}}{c-v} - \frac{3}{16} \frac{c \xi^{3/2}\w^{3/2}}{(c-v)^3} + O(\w^{5/2})\ , \\
    D_u &=& \frac{1}{\sqrt{\xi \w}} + \frac{1}{2}\frac{\sqrt{\xi \w}}{c+v} - \frac{3}{16}\frac{c \xi^{3/2}\w^{3/2}}{(c+v)^3} + O(\w^{5/2})\ , \\
    D_\pm &=& \sqrt{\frac{v/c}{c^2-v^2}} \bigg( \frac{\pm v \sqrt{c^2-v^2} - \I (c^2 - v^2)}{c \sqrt{\xi\w}} + \frac{\sqrt{\xi\w}}{2} \frac{\mp 2v \sqrt{c^2 - v^2} + \I (c^2 - 2v^2)}{(c^2 - v^2)} \\
    &+& \frac{c\xi^{3/2}\w^{3/2}}{16 v^2} \frac{\pm v(c^4 - 8c^2v^2 - 2v^4) - \I \sqrt{c^2-v^2}(c^4 - 6c^2v^2 + 2v^4)}{(c^2 - v^2)^{5/2}}\bigg) + O(\w^{5/2})\ ,  \nonumber    \label{eqn:ch4:subsonic-D-end}
\eea
\bea
    \label{eqn:ch4:subsonic-E-start} 
    E_v &=& -\frac{1}{\sqrt{\xi \w}} + \frac{1}{2}\frac{\sqrt{\xi \w}}{c-v} + \frac{3}{16} \frac{c \xi^{3/2}\w^{3/2}}{(c-v)^3} + O(\w^{5/2})\ , \\
    E_u &=& -\frac{1}{\sqrt{\xi \w}} + \frac{1}{2}\frac{\sqrt{\xi \w}}{c+v} + \frac{3}{16}\frac{c \xi^{3/2}\w^{3/2}}{(c+v)^3} + O(\w^{5/2})\ , \\
    E_\pm &=& \sqrt{\frac{v/c}{c^2-v^2}} \bigg( \frac{\mp v \sqrt{c^2-v^2} - \I (c^2 - v^2)}{c \sqrt{\xi\w}} + \frac{\sqrt{\xi\w}}{2} \frac{\mp 2v \sqrt{c^2 - v^2} - \I (c^2 - 2v^2)}{(c^2 - v^2)}  \\
    &+& \frac{c\xi^{3/2}\w^{3/2}}{16 v^2} \frac{\mp v(c^4 - 8c^2v^2 - 2v) - \I \sqrt{c^2-v^2}(c^4 - 6c^2v^2 + 2v^4)}{(c^2 - v^2)^{5/2}}\bigg) + O(\w^{5/2})\ , \nonumber
    \label{eqn:ch4:subsonic-E-end}
\eea
supersonic 
\bea
\label{eqn:ch4:supersonic-D-start} 
 D_v &=& \frac{1}{\sqrt{\xi \w}} + \frac{1}{2}\frac{\sqrt{\xi\w}}{c-v} - \frac{3}{16} \frac{c \xi^{3/2} \w^{3/2}}{(c-v)^3} + O(\w^{5/2})\ ,  \\
 D_u &=& -\frac{1}{\sqrt{\xi \w}} - \frac{1}{2}\frac{\sqrt{\xi\w}}{c+v} + \frac{3}{16} \frac{c \xi^{3/2} \w^{3/2}}{(c-v)^3} + O(\w^{5/2})\ ,  \\
 D_{3,4} &=& \frac{1}{\sqrt{2}} \frac{\sqrt{v^2 - c^2} \mp v}{(v^2 - c^2)^{3/4}} - \frac{c \xi \w}{2\sqrt{2}}\frac{v^2 + c^2 \mp v \sqrt{v^2 - c^2} }{(v^2 - c^2)^{9/4}} \\
 &+&\frac{c^2 \xi^2 \w^2}{16\sqrt{2}}\frac{10v^4 - 7c^2v^2 - 3c^4 \mp v\sqrt{v^2-c^2}(17c^2+10v^2)}{(v^2-c^2)^{17/4}} + O(\w^{5/2})\ , \nonumber
 \label{eqn:ch4:supersonic-D-end} 
\eea
\bea
\label{eqn:ch4:supersonic-E-start} 
 E_v &=& -\frac{1}{\sqrt{\xi \w}} + \frac{1}{2}\frac{\sqrt{\xi\w}}{c-v} + \frac{3}{16} \frac{c \xi^{3/2} \w^{3/2}}{(c-v)^3} + O(\w^{5/2})\ ,  \\
 E_u &=& \frac{1}{\sqrt{\xi \w}} - \frac{1}{2}\frac{\sqrt{\xi\w}}{c+v} - \frac{3}{16} \frac{c \xi^{3/2} \w^{3/2}}{(c-v)^3} + O(\w^{5/2})\ , \\ 
 E_{3,4} &=&  \frac{1}{\sqrt{2}}\frac{\sqrt{v^2-c^2}\pm v}{ \left(v^2-c^2\right)^{3/4}} + \frac{c \xi \w}{2 \sqrt{2}}\frac{c^2+v^2 \pm v \sqrt{v^2-c^2}}{ \left(v^2-c^2\right)^{9/4}} \\
 &+& \frac{c^2 \xi^2 \w^2}{16\sqrt{2}}\frac{10v^4 - 7c^2v^2 - 3c^4 \pm v\sqrt{v^2-c^2}(17c^2+10v^2)}{(v^2-c^2)^{17/4}} + O(\w^{5/2})\ , \nonumber
\label{eqn:ch4:supersonic-E-end} 
\eea
and sonic flows
\bea
 D_v &=& \frac{1}{\sqrt{\xi\w}} + \frac{\sqrt{\xi\w}}{4c} - \frac{\xi^{3/2} \w^{3/2}}{128 c^2} + O(\w^{5/2})\ , \\
D_u &=& \frac{1}{\sqrt{3}}\frac{1}{\sqrt{\xi \w}} + \frac{1}{\sqrt{3}} \frac{1}{c^{1/3} \xi^{1/6} \w^{1/6}} + \frac{1}{3\sqrt{3}}\frac{\xi^{1/6}\w^{1/6}}{c^{2/3}} - \frac{1}{12\sqrt{3}}\frac{\sqrt{\xi\w}}{c} + O(\w^{\frac{5}{6}}), \ \ \ \ \\
D_\pm &=& \frac{-1\pm\I \sqrt{3}}{\sqrt{6}}\frac{1}{\sqrt{\xi\w}} + \frac{-1\mp\I\sqrt{3}}{\sqrt{6}}\frac{1}{c^{1/3}\xi^{1/6}\w^{1/6}} + \frac{65 \mp 9\I\sqrt{3}}{48\sqrt{6}} \frac{\xi^{1/6}\w^{1/6}}{c^{2/3}} \\ 
    &+& \frac{1 \pm 17\I\sqrt{3}}{48\sqrt{6}} \frac{\sqrt{\xi\w}}{c} + O(\w^{5/6})\ , \nonumber
\eea
\bea
 E_v &=& -\frac{1}{\sqrt{\xi \w}}+\frac{\sqrt{\xi \w}}{4 c}+\frac{\xi ^{3/2} \w^{3/2}}{128 c^2}+O(\w^{5/2})\ , \\
 E_u &=& -\frac{1}{\sqrt{3}}\frac{1}{ \sqrt{\xi \omega }} + \frac{1}{\sqrt{3} }\frac{1}{c^{1/3} \xi^{1/6} \w^{1/6}} - \frac{1}{3 \sqrt{3}}\frac{\xi^{1/6} \w^{1/6}}{ c^{2/3}} - \frac{1}{12 \sqrt{3}} \frac{\sqrt{\xi } \sqrt{\omega }}{ c} +O(\omega^{\frac{5}{6}}),\ \ \ \ \ \  \\
E_\pm &=& -\frac{-1\pm\I \sqrt{3}}{\sqrt{6}}\frac{1}{\sqrt{\xi\w}} + \frac{-1\mp\I\sqrt{3}}{\sqrt{6}}\frac{1}{c^{1/3}\xi^{1/6}\w^{1/6}} - \frac{65 \mp 9\I\sqrt{3}}{48\sqrt{6}} \frac{\xi^{1/6}\w^{1/6}}{c^{2/3}} \\ 
   &+& \frac{1 \pm 17\I\sqrt{3}}{48\sqrt{6}} \frac{\sqrt{\xi\w}}{c} + O(\w^{5/6})\ . \nonumber
\eea

\section{Scattering matrix coefficients for the subsonic-supersonic model}
\label{appendixB}

The elements of the scattering matrix for the subsonic-supersonic model are as follows
\bea
    S_{ur,vr} &=& \frac{v + c_r}{v - c_r} + O(\w)\ , \\
    S_{vl,vr} &=& \frac{\sqrt{c_r}}{\sqrt{c_l}} \frac{(c_l - v)}{(c_r - v)} + O(\w)\ , \\
    S_{ul,vr} &=& \frac{\sqrt{c_r}}{\sqrt{c_l}} \frac{(c_l + v)}{(c_r - v)} + O(\w)\ , \\
    S_{ur,3l} &=& \frac{\sqrt{2 m c_r}}{\sqrt{\hbar\w}}\frac{(v^2 - c_l^2)^{3/4}}{(c_r^2 - c_l^2)} \frac{\sqrt{c_r + v}}{\sqrt{c_r - v}} (\sqrt{c_r^2 - v^2} + \I \sqrt{v^2 - c_l^2}) + O(\sqrt{\w})\ , \\
    S_{vl,3l} &=& \frac{\sqrt{m}}{\sqrt{2c_l \hbar \w}}\frac{(v^2 - c_l^2)^{3/4}}{(c_l + c_r)} \frac{\sqrt{c_r + v}}{\sqrt{c_r - v}} (\sqrt{c_r^2 - v^2} + \I \sqrt{v^2 - c_l^2}) + O(\sqrt{\w})\ , \\
    S_{ul,3l} &=& \frac{\sqrt{m}}{\sqrt{2c_l \hbar \w}}\frac{(v^2 - c_l^2)^{3/4}}{(c_l - c_r)} \frac{\sqrt{c_r + v}}{\sqrt{c_r - v}} (\sqrt{c_r^2 - v^2} + \I \sqrt{v^2 - c_l^2}) + O(\sqrt{\w})\ , \\
    S_{ur,4l} &=& \frac{\sqrt{2 m c_r}}{\sqrt{\hbar\w}}\frac{(v^2 - c_l^2)^{3/4}}{(c_r^2 - c_l^2)} \frac{\sqrt{c_r + v}}{\sqrt{c_r - v}} (\sqrt{c_r^2 - v^2} - \I \sqrt{v^2 - c_l^2}) + O(\sqrt{\w})\ , \label{uuu} \\
    S_{vl,4l} &=& \frac{\sqrt{m}}{\sqrt{2c_l \hbar \w}}\frac{(v^2 - c_l^2)^{3/4}}{(c_l + c_r)} \frac{\sqrt{c_r + v}}{\sqrt{c_r - v}} (\sqrt{c_r^2 - v^2} - \I \sqrt{v^2 - c_l^2}) + O(\sqrt{\w})\ , \label{vlvl}\\
    S_{ul,4l} &=& \frac{\sqrt{m}}{\sqrt{2c_l \hbar \w}}\frac{(v^2 - c_l^2)^{3/4}}{(c_l - c_r)} \frac{\sqrt{c_r + v}}{\sqrt{c_r - v}} (\sqrt{c_r^2 - v^2} - \I \sqrt{v^2 - c_l^2}) + O(\sqrt{\w})\ . \label{ulul}
\eea

\section{Numerical analysis for the three regions model (supersonic-sonic-subsonic)}
\label{appendixC}

In order to find the number of particles emitted in the three regions model, we wrote a Python module which is publicly available in \cite{daniel}. The module works as follows.

The inputs needed are the speeds of sound in the three (homogeneous) regions of the condensate, the (constant) speed of the flow, the mass and the density of the atoms forming the condensate, the coordinates of the discontinuities and the sample of frequencies used to solve the model. We typically chose the values $c_r = 2, c_s = 1 = -v, c_l = 0.5, m = 1, n = 1/(4\pi), \hbar=1$ and $x_1 = 0, x_2 = a$ for the coordinates of the discontinuities, with $a$ varying as shown in the main text. The sample of frequencies consisted of $10^{4}$ values ranging between $\w_{max} \simeq 0.295$ for the upper bound and $\w_{max}\times10^{-6}$ for the lower bound. 

The program 
first solves the dispersion relation (\ref{nove}) for each value of the frequency and identifies the different modes in the following way:
\begin{itemize}
    \item If it finds complex roots, then it identifies $k_v$ as the real negative solution, $k_u$ as the real positive solution, $k_+$ as the complex solution with positive imaginary part and $k_-$ as the complex solution with negative imaginary part.
    \item If it finds no complex roots, then it orders the roots ascendingly and identifies the modes following the relation $k_4 < k_u < k_v < k_3$ according to Fig. (\ref{fig:dispersion-supersonic}).
\end{itemize}

After this, the normalization coefficients (\ref{dieci}), (\ref{undici}) are calculated and the program identifies the characteristic of each region using the wavevectors $k$ calculated previously. This is done in order to find the appropriate dimensions of the scattering matrix and to identify which modes are ingoing, outgoing, decaying and growing (non-normalizable). For the thick horizon model, the model will correctly identify modes $vr, 3l, 4l$ as ingoing modes and $-r$ as a growing, non-normalizable mode.

Lastly, with the roots of the dispersion relation, normalization coefficients and knowing which modes are ingoing modes, the outgoing, decaying and sonic amplitudes are calculated for each ``in'' mode. In order to do this, we rewrite equations (\ref{pm}), (\ref{sm}) as
\begin{equation}
    \tilde{M} X = \tilde{N} A_{in}
\end{equation}
where $\tilde{M}$ is an $8\times8$ matrix, $X$ is an $8\times1$ matrix containing all the outgoing, decaying and sonic amplitudes, $\tilde{N}$ is an $8\times3$ matrix and $A_{in}$ is a $3\times1$ matrix containing the ingoing amplitudes.

The program calculates the entries of $\tilde{M}$ and $\tilde{N}$ and then calculates and returns the matrix $\tilde{M}^{-1} \tilde{N}$. The columns of this matrix are the outgoing, decaying and sonic amplitudes of each ``in'' mode, which is what we used to construct the scattering matrix in the main text.

\section{Analytical analysis for the three regions model (supersonic-sonic-subsonic)}
\label{appendixD}


We shall outline here the analytic procedure followed to construct all the amplitudes of the $4l$ `in' mode. 

The matching conditions at the left discontinuity surface ($x=0$) are
  \bea
        \begin{split}
    &A_{vl} D_{vl}  + A_{ul} D_{ul}  + D_{4l}  = A_{vs} D_{vs}  + A_{us} D_{us}  + A_{-s} D_{-s}  +A_{+s} D_{+s}, \label{left_matching_eq_1}
    \end{split} \\
    \begin{split}
    &k_{vl} A_{vl} D_{vl}  + k_{ul} A_{ul} D_{ul}  + k_{4l} D_{4l}  = k_{vs} A_{vs} D_{vs} + k_{us} A_{us} D_{us} + 
    k_{-s} A_{-s} D_{-s}  + k_{+s} A_{+s} D_{+s}\  ,
    \label{left_matching_eq_2}
    \end{split} \\
    \begin{split}
    &A_{vl} E_{vl}  + A_{ul} E_{ul}  + E_{4l}  = A_{vs} E_{vs}  + A_{us} E_{us}  + A_{-s} E_{-s}  +A_{+s} E_{+s} ,
    \label{left_matching_eq_3}    
    \end{split} \\  
    \begin{split}
    &k_{vl} A_{vl} E_{vl}  + k_{ul} A_{ul} E_{ul}  + k_{4l} E_{4l}  = k_{vs} A_{vs} E_{vs}  + k_{us} A_{us} E_{us}  + k_{-s} A_{-s} E_{-s}  + k_{+s} A_{+s} E_{+s} .
    \label{left_matching_eq_4}   
    \end{split}   
\eea
We combine the equations above in order to express $A_{-s}, A_{+s}$ as linear combinations of $A_{us}, A_{vs}$, finally getting
\begin{equation}
    \begin{cases}
A_{-s} = \frac{N_3^1 + N_3^2 A_{vs} + N_3^3 A_{us}}{D_3}, 
\\
A_{+s} = \frac{N_4^1 + N_4^2 A_{vs} + N_4^3 A_{us}}{D_4},
\end{cases}
\label{A_minus_s_A_plus_s}
\end{equation}
where $N_3^1, N_3^2, N_3^3, D_3, N_4^1, N_4^2, N_4^3, D_4$ are functions of Bogoliubov weights and momenta \cite{marco}.
The matching conditions at the right discontinuity surface $-$ i.e. at $x=a$ $-$ are
\bea
 &&   A_{vs} D_{vs} e^{i k_{vs} a} + A_{us} D_{us} e^{i k_{us} a} + A_{-s} D_{-s} e^{i k_{-s} a} +A_{+s} D_{+s} e^{i k_{+s} a}  
    = \, A_{ur} D_{ur} e^{i k_{ur} a} + A_{+r} D_{+r} e^{i k_{+r} a}, \ \ \ \ \ \ \ \ \label{cinque} \\
  &&  k_{vs} A_{vs} D_{vs} e^{i k_{vs} a} + k_{us} A_{us} D_{us} e^{i k_{us} a} + k_{-s} A_{-s} D_{-s} e^{i k_{-s} a} + k_{+s} A_{+s} D_{+s} e^{i k_{+s} a}
    = \,  \nonumber \\ && k_{ur} A_{ur} D_{ur} e^{i k_{ur} a} + k_{+r} A_{+r} D_{+r} e^{i k_{+r} a}, \label{dsei} \\
  &&  A_{vs} E_{vs} e^{i k_{vs} a} + A_{us} E_{us} e^{i k_{us} a} + A_{-s} E_{-s} e^{i k_{-s} a} +A_{+s} E_{+s} e^{i k_{+s} a}
    = \, A_{ur} E_{ur} e^{i k_{ur} a} + A_{+r} E_{+r} e^{i k_{+r} a},\ \ \ \ \ \  \\
 &&   k_{vs} A_{vs} E_{vs} e^{i k_{vs} a} + k_{us} A_{us} E_{us} e^{i k_{us} a} + k_{-s} A_{-s} E_{-s} e^{i k_{-s} a} + k_{+s} A_{+s} E_{+s} e^{i k_{+s} a} 
    = \, \nonumber \\ && k_{ur} A_{ur} E_{ur} e^{i k_{ur} a} + k_{+r} A_{+r} E_{+r} e^{i k_{+r} a}.
\eea
Therefore, substituting \eqref{A_minus_s_A_plus_s} into the right-side system yields
\bea
   && \alpha A_{vs} + \beta A_{us} + \gamma +\delta = A_{ur} D_{ur} e^{i k_{ur} a} + A_{+r} D_{+r} e^{i k_{+r} a}
\label{right_matching_eq_1}\ , \\
&&    \alpha' A_{vs} + \beta' A_{us} + \gamma' +\delta' = k_{ur} A_{ur} D_{ur} e^{i k_{ur} a} + k_{+r} A_{+r} D_{+r} e^{i k_{+r} a}
\label{right_matching_eq_2}\ , \\
   && \Tilde{\alpha} A_{vs} + \Tilde{\beta} A_{us} + \Tilde{\gamma} +\Tilde{\delta} = A_{ur} E_{ur} e^{i k_{ur} a} + A_{+r} E_{+r} e^{i k_{+r} a}
\label{right_matching_eq_3}\ , \\
   && \Tilde{\alpha}' A_{vs} + \Tilde{\beta}' A_{us} + \Tilde{\gamma}' +\Tilde{\delta}' = k_{ur} A_{ur} E_{ur} e^{i k_{ur} a} + k_{+r} A_{+r} E_{+r} e^{i k_{+r} a}
\label{right_matching_eq_4} \ ,
\eea
where again all the greek letters coefficients are functions of Bogoliubov weights and momenta \cite{marco}.
Notice that now we have 4 equations in 4 unknowns, thus giving a completely solvable system.
From a computational point of view, the procedure which ended up being effective indeed was to solve the previous system of equations with \textit{Wolfram Mathematica} directly. In particular, we first asked \textit{Wolfram Mathematica} to solve that system without defining any parameter explicitly, then we substituted the algebraic expressions of all of them into the implicit formula of $A_{ur}$ that we had got from \textit{Wolfram Mathematica}. The parameters have been series-expanded up to $O(\omega^2)$. Finally, at leading order we obtained
\begin{equation}A_{ur} (a) = 
    \sqrt{\frac{2c_l \xi_l c_r (c_r-c_s)\left(c_s^2-c_l^2\right)^{3/2}}{\omega(c_r+c_s)}}\frac{c_l \xi_l \left( \sqrt{c_r^2-c_s^2}-i\sqrt{c_s^2-c_l^2} \right)-i S_a}{\ S_a \left(a \sqrt{c_r^2-c_s^2} +c_l \xi_l \right) \sqrt{c_s^2-c_l^2}+c_l^2 \xi_l^2 \left(c_r^2-c_l^2\right)},
\label{eqn:ch6:aur4l-exact}
\end{equation}
where the \textit{sonic region damping} term $S_a$ has been defined as 
\be \label{df1} S_a \equiv 2 a  \sqrt{c_r^2-c_s^2} \sqrt{c_s^2-c_l^2}\ ,\ee 
such that $S_0 =0$. 
The case of an absent sonic region ($a=0$)
\begin{equation}
    A_{ur} (a=0) = \sqrt{\frac{2c_r (c_r-c_s)(c_s^2 - c_l^2)^{3/2}}{\omega c_l \xi_l (c_r+c_s)}} \frac{
    (\sqrt{c_r^2- c_s^2} - i\sqrt{c_s^2 - c_l^2})}{c_r^2 - c_l^2}
\end{equation}
consistently corresponds to the already known result in the two regions model \cite{capitolo-libro} . Furthermore, we verify the order $O(\frac{1}{\sqrt\omega})$ does vanish in the limit $a \to +\infty$, as we expect from the subsonic-sonic configuration.
The other subsonic amplitude is the decaying $A_{+r}$, which at leading order in $\omega$ reads
\begin{equation}
    A_{+r}(a)= \frac{i \, c_r \, (c_s^2-c_l^2)^{1/4}}{\sqrt{2} (c_r^2-c_s^2)} \sqrt{\frac{c_l \, \xi_l \,  \omega}{c_s}} e^{\frac{2 a \sqrt{c_r^2-c_s^2}}{c_l \, \xi_l}} \frac{S_a (a\sqrt{c_s^2-c_l^2}+i\, c_l \, \xi_l)\sqrt{c_r^2-c_s^2}-c_l^2 \, \xi_l^2 \, (c_r^2-c_l^2)}{S_a (a \sqrt{c_r^2-c_s^2}+c_l \, \xi_l)\sqrt{c_s^2-c_l^2}+c_l^2 \, \xi_l^2 \, (c_r^2-c_l^2)}.
\end{equation}
Again, in the absence of the sonic region ($a=0$) we recover the already known result in the two regions model, consistently \cite{capitolo-libro}:
\begin{equation}
    A_{+r}(a=0)= -\frac{i \, c_r \, (c_s^2-c_l^2)^{1/4}}{\sqrt{2} (c_r^2-c_s^2)} \sqrt{\frac{c_l \, \xi_l \,  \omega}{c_s}}.
\end{equation}
From the system of equations (\ref{right_matching_eq_1})-(\ref{right_matching_eq_4}) we can also compute the two propagating sonic amplitudes $A_{vs}$ and $A_{us}$. Here are their leading orders in $\omega$, respectively:
\begin{equation}
    A_{vs}=\frac{(c_s^2-c_l^2)^{3/4}(c_r-c_s)^{3/2}\sqrt{c_l \xi_l}}{\sqrt{2 c_s (c_r+c_s) \omega}}\frac{c_l \xi_l (\sqrt{c_r^2-c_s^2}-i\sqrt{c_s^2-c_l^2})-iS_a}{S_a(a\sqrt{c_r^2-c_s^2}+c_l \xi_l)\sqrt{c_s^2-c_l^2}+c_l^2 \xi_l^2 (c_r^2-c_l^2)},
\end{equation}
\begin{equation}
    A_{us}=\frac{(c_s^2-c_l^2)^{3/4} \sqrt{c_r^2-c_s^2} \sqrt{c_l \xi_l}}{\sqrt{6 c_s \omega}}\frac{c_l \xi_l (\sqrt{c_r^2-c_s^2}-i\sqrt{c_s^2-c_l^2})-iS_a}{S_a(a\sqrt{c_r^2-c_s^2}+c_l \xi_l)\sqrt{c_s^2-c_l^2}+c_l^2 \xi_l^2 (c_r^2-c_l^2)}.
\end{equation}
Upon (\ref{A_minus_s_A_plus_s}), it is possible to calculate the growing and decaying sonic amplitudes $A_{-s}$ and $A_{+s}$, which at leading order in $\omega$ appear as
\begin{equation}
    A_{-s}=\frac{(-\sqrt{3}+3i)(c_s^2-c_l^2)^{3/4} \sqrt{c_r^2-c_s^2} \sqrt{c_l \xi_l}}{12\sqrt{c_s \omega}}\frac{c_l \xi_l (\sqrt{c_r^2-c_s^2}-i\sqrt{c_s^2-c_l^2})-iS_a}{S_a(a\sqrt{c_r^2-c_s^2}+c_l \xi_l)\sqrt{c_s^2-c_l^2}+c_l^2 \xi_l^2 (c_r^2-c_l^2)},
\end{equation}
\begin{equation}
    A_{+s}=\frac{-(\sqrt{3}+3i)(c_s^2-c_l^2)^{3/4} \sqrt{c_r^2-c_s^2} \sqrt{c_l \xi_l}}{12\sqrt{c_s \omega}}\frac{c_l \xi_l (\sqrt{c_r^2-c_s^2}-i\sqrt{c_s^2-c_l^2})-iS_a}{S_a(a\sqrt{c_r^2-c_s^2}+c_l \xi_l)\sqrt{c_s^2-c_l^2}+c_l^2 \xi_l^2 (c_r^2-c_l^2)}.
\end{equation}
Since we have solved for the 6 amplitudes $A_{ur}, A_{+r}, A_{vs}, A_{us}, A_{-s}, A_{+s}$, in order to calculate the two remaining amplitudes in the supersonic region we just need to consider two equations properly combining $A_{vl}$ and $A_{ul}$ out of the matching conditions \eqref{left_matching_eq_1}-\eqref{left_matching_eq_4} at the left discontinuity. We are searching for leading order in $\omega$ solutions, therefore we shall first notice the left-hand side of \eqref{left_matching_eq_1} and the right-hand side of \eqref{cinque} equate at leading order for any value of $a$
\begin{equation}
    A_{vl}-A_{ul} = \sqrt{\frac{c_r}{c_l}} A_{ur}\ .
\label{eq_Sandro_1}
\end{equation}
Then, we shall take the difference between \eqref{left_matching_eq_2} and \eqref{dsei} at leading order into account, thus yielding
\begin{equation}
\begin{split}
    &-\frac{1}{\sqrt{2 \xi_l}}\left(\frac{A_{ul} \sqrt{\omega}}{c_l-c_s}
    +\frac{A_{vl} \sqrt{\omega}}{c_l+c_s}\right)-\frac{A_{ur} \sqrt{\omega}}{\sqrt{2 \xi_r} (c_r-c_s)}+\frac{c_s-\sqrt{c_s^2-c_l^2}}{c_l \xi_l (c_s^2-c_l^2)^{1/4}}-\frac{2 i \sqrt{c_r^2-c_s^2} A_{+r} D_{+r} e^{-\frac{2 a \sqrt{c_r^2-c_s^2}}{c_r \xi_r}}}{c_r \xi_r}
    \\
    &=k_{us} D_{us} A_{us}(1 - e^{i k_u^s a}) + k_{+s} D_{+s} A_{+s}(1 - e^{i k_{+s} a}) + k_{-s} D_{-s} A_{-s}(1 - e^{ik_{-s} a}).
\label{eq_Sandro_2}
\end{split}
\end{equation}
In the right-hand side of the latter equation we did not include the term $k_v^s D_v^s A_v^s (1-e^{i k_v^s a})$ because its leading order is $O(\omega)$, while the leading order of the left-hand side is $O(1)$. In addition, the right-hand side of \eqref{eq_Sandro_2} contains internal cancellations, for each term has $O(\frac{1}{\omega^{1/3}})$ as leading order. To compute $A_{vl}$ and $A_{ul}$ by \eqref{eq_Sandro_1} and \eqref{eq_Sandro_2} we consider also the next-to-leading order results, eventually getting
\begin{equation}\label{aul}
\begin{split}
    A_{ul}(a) =
    &\sqrt{\frac{\xi_l}{2 \omega} }
    \frac{1}{\sqrt{c_r^2-c_s^2}
    \left[
    S_a \sqrt{c_s^2-c_l^2} \left(a \,  \sqrt{c_r^2-c_s^2}  
    +c_l \xi_l\right)
    + c_l^2 \xi_l^2 \left(c_r^2-c_l^2 \right) 
    \right]^2
    }
    \\
    &\times
    \left[
    -(c_r +c_l)(c_r-c_s)(c_s^2-c_l^2)^{3/4}
    \left(c_l^3 \xi_l^3 (c_r^2-c_l^2) \left(\sqrt{c_r^2-c_s^2} -i \sqrt{c_s^2-c_l^2} \right)
    \right.
    \right.
    \\
    &-a \sqrt{c_s^2-c_l^2}
    \left\{
    S_a \sqrt{c_r^2-c_s^2}
    \left[
    iS_a -c_l \xi_l 
    \left(
    \sqrt{c_r^2-c_s^2}-3 i \sqrt{c_s^2-c_l^2} 
    \right)
    \right]
    \right.
    \\
    &\left.
    \left.
    \left.
    +2 i c_l^2 \xi_l^2
    \left[
    c_r^2 \left( \sqrt{c_r^2-c_s^2} +i \sqrt{c_s^2-c_l^2} \right)
    +c_s^2 \left( \sqrt{c_r^2-c_s^2} -i \sqrt{c_s^2-c_l^2} \right)
    -2 c_l^2 \sqrt{c_r^2-c_s^2}
    \right]\right\}\right)\right],
\end{split}
\end{equation}
\begin{equation}\label{avl}
\begin{split}
    A_{vl}(a) =
    &\sqrt{\frac{\xi_l}{2 \omega} }
    \frac{1}{\sqrt{c_r^2-c_s^2}
    \left[
    S_a \sqrt{c_s^2-c_l^2} \left(a \,  \sqrt{c_r^2-c_s^2}  
    +c_l \xi_l\right)
    + c_l^2 \xi_l^2 \left(c_r^2-c_l^2 \right) 
    \right]^2
    }
    \\
    &\times
    \left[
    (c_r-c_l)(c_r-c_s)(c_s^2-c_l^2)^{3/4}
    \left(c_l^3 \xi_l^3 (c_r^2-c_l^2) \left(\sqrt{c_r^2-c_s^2} -i \sqrt{c_s^2-c_l^2} \right)
    \right.
    \right.
    \\
    &-a \sqrt{c_s^2-c_l^2}
    \left\{
    S_a \sqrt{c_r^2-c_s^2}
    \left[
    iS_a -c_l \xi_l 
    \left(
    \sqrt{c_r^2-c_s^2}-3 i \sqrt{c_s^2-c_l^2} 
    \right)
    \right]
    \right.
    \\
    &\left.
    \left.
    \left.
    +2 i c_l^2 \xi_l^2
    \left[
    c_r^2 \left( \sqrt{c_r^2-c_s^2} +i \sqrt{c_s^2-c_l^2} \right)
    +c_s^2 \left( \sqrt{c_r^2-c_s^2} -i \sqrt{c_s^2-c_l^2} \right)
    -2 c_l^2 \sqrt{c_r^2-c_s^2}
    \right]\right\}\right)\right],
\end{split}
\end{equation}
where the damping factor $S(a)$ is given in (\ref{df1}).
We observe that not only do these expressions recover the already known results of Appendix (\ref{appendixB}) in the supersonic-subsonic two regions model when evaluating $a=0$, but they also vanish in the limit $a \to +\infty$ consistently with our results of subsection (\ref{ss}). Finally, our leading in $\omega$ analytical results for $A_{ur}(a), A_{ul}(a), A_{vl}(a)$, Eqs. (\ref{eqn:ch6:aur4l-exact}) (\ref{aul}), (\ref{avl}),  are consistent with the pseudo unitarity condition for the  $4l$ `in' mode (see Eq. (1.95) of \cite{capitolo-libro}) \be \label{pun} |A_{ur}(a)|^2-|A_{ul}(a)|^2+|A_{vl}(a)|^2=-1\ .\ee 

\end{appendix}

\end{document}